\documentclass[aps,fleqn,twocolumn,secnumarabic,nobalancelastpage,amsmath,
amssymb,nofootinbib]{revtex4-2}
\usepackage{dcolumn}
\usepackage{bm}
\usepackage[latin1]{inputenc}
\usepackage[active]{srcltx}
\usepackage{nicefrac}
\usepackage[plainpages=false,pdfpagelabels,colorlinks=true,linkcolor=red,urlcolor=blue,
citecolor=blue,pdftitle={prb-therm9},pdfauthor={},pdfdisplaydoctitle=true]{hyperref}
\usepackage{ae,aecompl}
\usepackage{graphicx}
\usepackage{amsmath}
\usepackage{color}
\usepackage{braket}
\usepackage{verbatim}

\begin{document}

\title{Spin-orbit coupling effects over thermoelectric transport properties in quantum dots}

\author{M.~A.~Manya$^{1}$}
\author{G.~B.~Martins$^{2}$}
\author{M.~S.~Figueira$^{1}$}

\email[corresponding author:]{figueira7255@gmail.com}
\affiliation{$^{1}$Instituto de F\'{\i}sica, Universidade Federal Fluminense, Av. Litor\^anea s/N, CEP: 24210-340, Niter\'oi, RJ, Brasil}

\affiliation{$^{2}$Instituto de F\'{\i}sica, Universidade Federal de Uberl\^andia, Uberl\^andia, Minas Gerais, 38400-902, Brazil}

\date{\today}

\begin{abstract}
	We study the effects caused by Rashba and Dresselhaus 
	spin-orbit coupling  over the thermoelectric transport properties 
	of a single-electron transistor, viz., a quantum dot connected to 
	one-dimensional leads. Using linear response 
	theory and employing the numerical renormalization 
	group method, we calculate the thermopower, electrical and thermal 
	conductances, dimensionless thermoelectric figure of merit, 
	and study the Wiedemann-Franz law, showing their temperature maps. 
	Our results for all those properties indicate that spin-orbit coupling drives the 
	system into the Kondo regime. We  show that the thermoelectric transport 
	properties, in the presence of spin-orbit coupling, 
	obey the expected universality of the Kondo strong coupling fixed point. 
	In addition, our results show a notable increase in the thermoelectric figure of 
	merit, caused by the spin-orbit coupling in the one-dimensional quantum dot leads. 
\end{abstract}


\maketitle

\section{Introduction}
\label{sec1}

The discovery of the Seebeck and Peltier effects on junctions of different metals
at the beginning of the $19^{th}$ century gave rise to a part of thermal science
named ``Thermoelectricity"~\cite{Rosa16}. The
Seebeck effect is the voltage bias that develops when two
different metals are joined together (forming a thermocouple), with their junctions
maintained at different temperatures. Ten years after Seebeck's
discovery, Peltier observed that heat is either absorbed or rejected 
when an electric current flows through a Seebeck device,
depending on the current direction along the circuit. Nowadays, 
Peltier and Seebeck's effects constitute the basis for many 
thermoelectric (TE) refrigeration and TE power generation devices, 
respectively~\cite{Tritt2002,Shakouri2011,Tritt2011}.

Modern high-performance thermoelectric materials have their maximum
dimensionless TE figure of merit, $ZT$, in the interval $ZT\simeq 1$ 
to $2.5$ (see Fig.~2 of Ref.~\cite{Jian17}), which is well below the 
Carnot cycle efficiency~\cite{Benenti2017}. However, from the technological 
perspective,  a considerable $ZT$ value in a wide range of temperatures 
is preferable to a localized $ZT$ peak. On the other hand, $ZT$ must 
attain values between $3$ and $4$ to compete with other energy-generation 
processes; that is the reason why TE generators (and refrigerators) are 
not part of our daily life. There are some niches in particular fields, 
like the satellite and aerospace industry, where the advantages of not 
having movable parts and not requiring maintenance overcome their low 
efficiency and higher costs~\cite{Jian17}. One example is the radioisotope 
TE generator~\cite{Bourouis20}, a nuclear electric generator that employs 
a radioactive-atom's natural decay (usually Plutonium Dioxide, ${^{238}}PuO_{2}$), 
to convert, via the Seebeck effect, the heat released by the disintegrated 
atoms into electricity.

With the improvement of experimental nanotechnology techniques, new possibilities 
for increasing $ZT$ arise, mainly due to the level of quantization and Coulomb 
interaction present in nanoscopic devices. Some promising compounds are 
topological insulators
(TIs), as well as Weyl and Dirac semi-metals, characterized by nontrivial
topological order. Mostly due to spin-orbit interaction~\cite{Manchon2015},
a novel characteristic of TIs is that besides
having a conventional semiconductor bulk band structure,
they also exhibit topological surface conducting-states~\cite{Martins2021}.
Some of the best TE materials are also three-dimensional topological insulators, such as
$Bi_{2}Te_{3}$, $Bi_{2}Se_{3}$, $Sb_{2}Te_{3}$ and $FeSb_{2}$~\cite{Xu2017,Xu2020,FeSb22012,gooth2018}.

In an earlier work~\cite{Ramos2014}, one of the authors has addressed the TE 
properties of a single-electron transistor (SET) constituted of a 
correlated quantum dot (QD) embedded into conducting leads, as represented schematically  
in Fig.~\ref{fig01}. Here, employing the numerical renormalization group 
(NRG) method~\cite{bulla2008numerical,Hewson}, we study the 
effect of conduction band spin-orbit coupling (SOC) 
over an SET's TE transport properties, viz., electrical and thermal 
conductances, thermopower, Wiedemann-Franz law, and the dimensionless 
TE figure of merit. As main results, we show that SOC drives the system 
into the Kondo regime, where the universality of those properties is 
satisfied~\cite{Seridonio_2009,Seridonio1_2009,Seridonio2_2009,Seridonio_2010,
costi2010thermoelectric,Roberto2021}, although we have found the 
interesting result that the universality of the thermopower is better 
fulfilled at the intermediate valence regime than at the Kondo regime. 
More importantly, we show that 
SOC causes a notable increase in the dimensionless figure of merit of an SET. 
Our analysis is done at low enough temperatures 
to warrant the neglect of the phononic contribution to the SET TE properties~\cite{Zianni2010}. 
In addition, it is worth noting that there are also ways of decreasing 
the detrimental influence of phonons in the thermal efficiency of SETs by, for example, 
alloying the SET tunnel barriers to scatter phonons 
away from the QD~\cite{Ramos2014,Zianni2010,Vineis2010}. 

A strong motivation for studying SET physics is that this 
system is the experimental realization of the 
single impurity Anderson model (SIAM)~\cite{Anderson1961,Hewson} for finite
electronic correlation $U$. The SIAM was experimentally realized 
by Mark Kastner and Goldhaber-Gordon~\cite{Goldhaber2001}, when complete 
control over all the model parameters was achieved. They measured the electric 
conductance of a QD and showed its universal character. In the last years, the 
interest in the TE properties of QDs has greatly increased, 
yielding several papers, originating from theoretical 
\cite{Yoshida2009,costi2010thermoelectric,Hershfield13,Donsa14,Talbo2017,Costi20191,
Costi20192,Thierschmann2019,Eckern2020} as well as experimental groups 
\cite{Heremans2004,Scheibner2007,Hoffmann2009,Dutta2017,Hartman2018,
Artis2018,Dutta2019}. Recent reviews can be found in 
Refs.~\cite{Jian17,Rosa16,Benenti2017}.

Since a few decades ago, SOC has had a major impact in the development
of new information technologies~\cite{DasSarma2004,Bader2010}, especially after the discovery of TIs~\cite{Hasan2010}. This has
intensified studies of systems where SOC is determinant in
providing access to the spin degree of freedom~\cite{Winkler2003,Manchon2015}.
Furthermore, electron correlations and SOC may combine to produce new emergent
behavior~\cite{Pesin2010,Balents2014,Rau2016,Schaffer2016}, as e.g.
in Iridates, $\rm{Sr_2IrO_4}$~\cite{Kim2008} and TIs~\cite{Allerdt2017}. 
Another class of materials where SOC could play an important role are the topological Kondo insulators, of
which $SmB_{6}$ is the first example. The strong correlation between localized 
$4f$ states and the conduction $5d$ band gives rise to a bulk insulating state 
at low temperatures while the surface remains metallic. Recent experimental 
results points out that the surface states around the $X$ point of the 
Brillouin zone can be described by a combination of Rashba- and Dresselhaus-like 
SOC \cite{Xu2014,Zhu_2016,Li2020,Ryu2021}. As mentioned above, in this work, 
the authors use the NRG method~\cite{bulla2008numerical} to
study in an unbiased manner the TE properties of a QD in the Kondo regime under the
influence of SOC. Previous works involving Kondo and SOC can be
found in Refs.~\cite{Meir1994,Malecki2007,Zitko2011,Zarea2012,Mastrogiuseppe2014,
Wong2016,Chen2016,Sousa2016,Chen2017,Martins2020}. A description of 
the results obtained until recently, regarding the influence of SOC in the Kondo effect, 
can be found in Ref.~\cite{Martins2020}. To the best of our knowledge, 
the present work is the first to discuss the effect of SOC over the TE 
properties of an SET in the Kondo, intermediate valence, and 
empty-orbital regimes. 

We organize the paper in the following way: In Sec.~\ref{sec2}, we 
introduce the SIAM in the presence of Rashba~\cite{Bychkov1984} 
and Dresselhaus~\cite{Dresselhaus1955} conduction band SOC. 
In Sec.~\ref{sec4}, we present the formalism employed 
in the calculation of the TE properties. In Sec.~\ref{sec5}, 
we discuss the SOC effects over the QD local density 
of states (LDOS), $\rho_d(\omega)$, and over the conduction band 
DOS, $\rho_c(\omega)$. In Sec.~\ref{sec6}, we present the temperature maps of the TE 
properties. In Sec.~\ref{sec7} we discuss the universality of the TE 
properties under the effect of SOC. In Sec.~\ref{sec8}, we present 
a summary of the results and  their physical consequences. 

\section{Model and Theory}
\label{sec2}

\begin{figure}[htb]
\includegraphics[clip,width=0.45\textwidth,angle=0.]{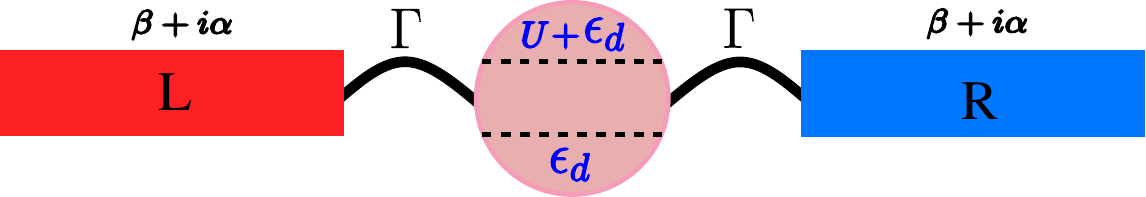}
\caption{Schematic representation of an SET, a correlated QD, 
	with $\epsilon_d$ and $\epsilon_d+U$ energy levels, 
	and Coulomb repulsion $U$. The QD is symmetrically coupled, 
	with strength $\Gamma$, to left and right 
	uncorrelated conducting leads. The one-dimensional leads 
	are subjected to Rashba ($\alpha$) and Dresselhaus ($\beta$) SOC.}
\label{fig01}
\end{figure}

The SET studied in this work is represented schematically in 
Fig.~\ref{fig01}. It is constituted by a correlated QD 
immersed into one-dimensional (1D) conducting leads that exhibit both 
Rashba~\cite{Bychkov1984} and Dresselhaus~\cite{Dresselhaus1955} SOC. 
The total Hamiltonian is given by $H = H_{leads} + H_{QD} + H_{hyb}$, where
\begin{eqnarray}
	H_{leads} &=&-2t\sum_{k\sigma s }[cos(ka)-\mu]
	c^{\dagger}_{k\sigma,s}c_{k\sigma,s} \nonumber \\
	&-&2\sum_{ks} \sin(ka)(\gamma c_{k\uparrow,s}^{\dagger}c_{k\downarrow,s} 
	+ h.c.),
  \label{Hleads}
\end{eqnarray}
\begin{eqnarray}	
H_{QD} = \sum_{\sigma} \epsilon_dn_{d\sigma} + Un_{d\uparrow}n_{d\downarrow}, 
\label{HQD}
\end{eqnarray}
\begin{eqnarray}
  H_{hyb}= \sum_{k\sigma s}V_{k\sigma,s}(c_{k\sigma,s}^{\dagger} d_{\sigma} + d_{\sigma}^{\dagger} c_{k\sigma,s}) .
	\label{Hhib}
\end{eqnarray}
In the equations above, $c_{k\sigma,s}^{\dagger}$ ($c_{k\sigma,s}$) creates 
(annihilates) an electron with momentum $k$ and spin $\sigma=\uparrow,\downarrow$ 
in the $s=L,R$ lead, while $d_{\sigma}^{\dagger}$ ($d_{\sigma}$) creates 
(annihilates) an electron with spin $\sigma=\uparrow,\downarrow$ in the QD, 
being $n_d=d_{\sigma}^{\dagger}d_{\sigma}$ the number operator for the $\epsilon_d$ 
QD active level. More 
specifically, $H_{leads}$ represents the leads, modeled by a 1D tight-binding 
approximation with nearest-neighbor hopping $t$, lattice parameter $a = 1$, 
and chemical potential $\mu$. Rashba ($\alpha$) and Dresselhaus ($\beta$) SOC parameters 
are taken into account through $\gamma = \beta + i \alpha$. 
$H_{QD}$ describes the QD, characterized by a localized bare level $\epsilon_d$ 
and a local Coulomb repulsion $U$, while $H_{hyb}$ accounts for the hybridization 
between electrons in the leads and the QD. In the following, we consider that 
the matrix elements of the hybridization 
$V_{k\sigma,s}$, originating from the coupling between the conducting leads and the QD 
electron's wave function, are $k$-, spin-, and lead-independent, i.e., $V_{k\sigma,s} = V $. 
Finally, we assume that $\epsilon_d$ can be varied by application of a voltage to a 
metallic back gate. 

As shown in previous works~\cite{Meir1994,Malecki2007,Zitko2011,Zarea2012,Mastrogiuseppe2014,
Wong2016,Chen2016,Sousa2016,Chen2017,Martins2020}, the inclusion 
of SOC in the conduction band results in a broken spin SU(2) 
symmetry. We may define a helicity operator $\hat{h}$ with eigenstates 
$|k\nu\rangle$, such that $[\hat{h},H_{leads}] = 0$. However, unfortunately, 
$\nu=\pm 1$ is not a good quantum number of the whole system (leads+QD), because the helicity is not 
defined for the QD.  Thus, it is convenient, \emph{in 1D}, 
to make a spin rotation to partially recover the SU(2) symmetry~\cite{Martins2020}. 
Therefore, we choose a new spin 
basis $S_r$, in which both the impurity and the conduction spins are 
projected along an axis $\hat{r}$ that points along the SOC effective magnetic field, 
whose orientation depends on the SOC parameter $\gamma$~\cite{Martins2020}. 
The basis transformations for impurity and SOC conduction band are 
$d^{\dagger}_{\sigma_r} = \nicefrac{1}{2}(d^{\dagger}_{\uparrow} + \sigma_re^{i\phi}d^{\dagger}_{\downarrow}) $ 
and $c^{\dagger}_{k\sigma_r} = \nicefrac{1}{2}(c^{\dagger}_{k\uparrow} + 
\sigma_re^{i\phi}c^{\dagger}_{k\downarrow}) $, 
respectively, where $\sigma_r = \pm$ indicates spins $\uparrow$ ($+$) and $\downarrow$ ($-$) 
along the $\hat{r}$ direction. In addition, $\phi = \nicefrac{\alpha}{\beta}$. 
In this new basis, the Hamiltonian can be rewritten as

\begin{eqnarray}
\label{eq:Hdiag}
	H&=&\sum _{k,\sigma_r}\epsilon_{k\sigma_r}c_{k\sigma_r}^{\dagger}c_{k\sigma_r}  \nonumber \\
	&+& \sum_{\sigma_r} \epsilon_dn_{d\sigma_r} + Un_{d\uparrow_r}n_{d\downarrow_r} \nonumber \\
	&+&\sum _{k,\sigma_r}V\left( c_{k\sigma_r}^{\dagger}d_{\sigma _{r}} +d_{\sigma _{r}}^{\dagger}c_{k\sigma_r}\right),
\end{eqnarray}
where $n_{d\sigma _{r}}=d_{\sigma _{r}}^{\dagger}d_{\sigma _{r}}$ is 
the QD number operator, $c_{k\sigma_r}^{\dagger}(c_{k\sigma_r})$ 
creates (annihilates) an electron in the Fermi sea with momentum $k$ and 
spin $\sigma_r$. 
Note that we have removed the $s=L,R$ subindex, since we are assuming that the QD couples 
only to the symmetric combination of the $L$ and $R$ leads. 
Finally, the dispersion relation $\epsilon_{k\sigma_r}$ is given by

\begin{equation}
\epsilon_{k\sigma_r}=-2\sqrt{t^{2}+\vert\gamma\vert ^{2}}\cos \left( k-\sigma_r\varphi\right) + \mu ,
\label{eq:eksigmar}
\end{equation}
where $\varphi = \tan^{-1} \nicefrac{| \gamma |}{t}$. 
As shown in Ref.~\cite{Martins2020}, in 1D the SOC influence over the Kondo effect is to 
cause a renormalization of the zero-SOC half-bandwidth $D$ and the 
QD-band hybridization, at the Fermi energy, $\Gamma = \Delta(0)$, to new $\gamma$-dependent values 
${D}_{\gamma}=2\sqrt{t^{2}+\vert\gamma \vert^{2}}$~\cite{note0} and 
\begin{equation}
\label{eq:Deltaren}
	{\Gamma}_{\gamma}={\Delta}_{\gamma}(0)= \frac{V^{2}}{2\sqrt{t^{2}+\vert\gamma\vert ^{2}}} .
\end{equation}
Thus, the renormalized finite-SOC SIAM Kondo temperature ${T}_{K\gamma}$, in 
the wide-band limit, can be written as~\cite{Martins2020}
\begin{equation}
\label{eq:hald1}
	{T}_{K\gamma} =0.364 \left(\frac{2{\Gamma}_{\gamma} U}{\pi}\right)^{\frac{1}{2}}
	\exp\left[\frac{\pi \epsilon_d\left(\epsilon_d+U \right)}{2{\Gamma}_{\gamma} U}\right].
\end{equation} 

On the other hand, Friedel's sum rule~\cite{Langreth66} gives 
a relationship, at $T=0$, between the extra state (induced below the Fermi level by a 
scattering center) and the phase shift at the chemical potential $\eta_{\sigma }(\mu )$, 
obtained by the transference matrix $T_{d,k\sigma_r}(z)=V_k^{2}G_{d,k\sigma_r}^{imp}(z)$, 
where $V_{k}$ is the scattering potential. For the SIAM, the extra states induced are 
given by the occupation number $n_{d,\sigma_r}$ of the QD localized state, and the 
scattering potential is the hybridization that affects the conduction electrons. 
Thus, Friedel's sum rule for the SIAM can be written as~\cite{Kang2001}
\begin{equation}\label{fried}
	\rho _{d\sigma }(\mu )=\frac{sin^{2} \left(\nicefrac{\pi n_{d\sigma }}{2}\right)}{{\Gamma}_{\gamma} \pi }, 
\end{equation}%
\noindent where $\rho _{d\sigma }(\mu )$ is the LDOS of the QD  
level at the chemical potential. In the Kondo regime, 
Eq.~\eqref{fried} implies that a suppression of ${\Gamma_{\gamma}}$ 
leads to an enhancement of the so-called Kondo peak at the chemical potential and the consequent 
suppression of ${T}_{K\gamma}$, due to the concomitant narrowing 
of the Kondo peak. Since SOC suppresses ${\Gamma_{\gamma}}$ [see Eq.\eqref{eq:Deltaren}], 
it results that SOC suppresses ${T}_{K\gamma}$ [this can also be concluded from Eq.~\eqref{eq:hald1}]. 

To perform the NRG calculations, we use the open-source NRG code 
NRG-Ljubljana~\cite{ZITKO20112259}. 
The NRG parameters used were $\Lambda = 2$, Wilson chain length $68$, 
keeping 20000 states. In addition, for all our calculations we use 
a zero-SOC hybridization function at the Fermi 
level ${\Gamma}_0 = 0.007$~\cite{note2}.

\section{Thermoelectric properties}
\label{sec4}

To calculate the TE transport properties of a QD in 
a steady-state condition, we apply a small external bias voltage 
$\Delta V=V_{L}-V_{R}$ and a small temperature difference $\Delta T=T_{L}-T_{R}$ 
between the left (hot) and the right (cold) leads. In linear response theory, 
a current $J_{\alpha}$ will flow through the system under the action of a temperature 
gradient $\vec{\nabla} T$ and/or an electric field $\vec{E}=-\vec{\nabla} V$, 
where $\alpha=e$ indicates a charge current $J_e$, while $\alpha=Q$ indicates a 
heat current $J_Q$. The TE properties calculations 
follow standard textbooks~\cite{Mahan, Ziman}. The electrical and thermal conductances, 
$G(T)$ and $K_e(T)$, respectively, as well as the thermopower $S(T)$ 
(Seebeck coefficient) are given by~\cite{note1}
\begin{equation}
G(T)=-\lim_{\Delta V \rightarrow 0} \left(J_{e}/\Delta V\right)_{\mid_{\Delta T=0}}=e^{2}\mathcal{L}_{0}(T) ,
\label{G}
\end{equation}
\begin{eqnarray}
K_e(T)&=&-\lim_{\Delta T \rightarrow 0} \left(J_{Q}/\Delta T\right)_{\mid_{J_{e}=0}} \\ \nonumber
	&=& \frac{1}{T}\left(\mathcal{L}_{2}(T)-\frac{\mathcal{L}_{1}^{2}(T)}{\mathcal{L}_{0}(T)}\right), 
\label{K}
\end{eqnarray}
\begin{equation}
S(T)=\lim_{\Delta T \rightarrow 0} \left(\Delta V/\Delta T\right)_{\mid_{J_{e}=0}}=\left(\frac{-1}{eT}\right) \frac{\mathcal{L}_{1}(T)}{\mathcal{L}_{0}(T)} ,
\label{S}
\end{equation}
where, to calculate the transport coefficients $\mathcal{L}_{0}(T)$, $\mathcal{L}_{1}(T)$, and 
$\mathcal{L}_{2}(T)$, we follow Ref.~\cite{Dong_02}, 
where expressions for the particle current and thermal flux, for a QD, were derived within 
the framework of Keldysh non-equilibrium Green's functions. Thus, the 
TE transport coefficients were obtained in the presence of 
temperature and voltage gradients, with the Onsager relations 
automatically satisfied, in the linear regime. The TE transport coefficients 
(for $n=0,1,2$) consistent with the general TE formulas derived above are given by

\begin{equation} \label{L11} 
\mathcal{L}_{n}(T)=\frac{2}{h}\int{\left(-\frac{\partial
f(\omega,T) }{\partial \omega} \right) \omega^{n}\tau(\omega,T) d\omega} ,
\end{equation}
\noindent where $\tau(\omega,T)$ is the transmittance for electrons with 
energy $\epsilon=\hbar\omega$ and temperature $T$, while $f(\omega,T)$ is 
the Fermi-Dirac distribution function.

For ordinary metals, the Wiedemann-Franz law states that 
the ratio between the electronic contribution to the thermal 
conductance $K_e(T)$ and the 
product of temperature $T$ and electrical conductance $G(T)$,  
\begin{equation}\label{LL}
L=\frac{K_e(T)}{TG(T)} ,
\end{equation}
is independent of temperature and takes a universal value given by 
the Lorenz number $L_{o}=(\nicefrac{\pi^{2}}{3})(\nicefrac{k_{B}}{e})^{2}$, where $k_{B}$ is 
the Boltzmann constant and $-e$ is the electron charge. Note that, in the calculations that follow, 
we present the Wiedemann-Franz law in units of $L_{o}$, 
\begin{equation}\label{WFL}
WF=\frac{L}{L_{o}} ,
\end{equation}
so that it is easy to spot deviations from what is expected for ordinary metals, i.e., $WF=1$.  

The dimensionless TE figure of merit $ZT$, which measures the 
efficiency of materials or devices to be employed as thermopower 
generators or cooling systems, is defined by
\begin{equation}\label{ZT}
	ZT=S^{2}TG(T)/K_e(T) .
\end{equation}
In ordinary metals, both $G(T)$ and $K_e(T)$ are related to the 
same electronic scattering processes, with only weak energy dependence, 
and satisfying the Wiedemann-Franz law, which is the main reason why metals 
show lower  $ZT$ values. The conditions that a  device must fulfill to 
produce a high $ZT$ value were discussed in Ref.~\cite{Sofo}, where the 
authors showed that a narrow energy distribution of the carriers was needed 
to produce a large value of $ZT$. 

Employing Eqs.~\eqref{LL} to \eqref{ZT}, we can write $ZT$ as a function 
of the thermopower $S$ and the Wiedemann-Franz law $WF$ [expressed in 
units of $L_o$, as in Eq.~\eqref{WFL}]
\begin{equation}\label{ZT1}
ZT=\frac{S^{2}}{WF} .
\end{equation}
This equation shows that violations of the Wiedemann-Franz law 
can lead to an effective increase of $ZT$ in regions where $WF<1$, 
as long as there is no concomitant decrease in the thermopower $S$.

The behavior of the TE coefficients is governed by the transmittance 
$\tau(\omega, T)$ [see Eq.~\eqref{L11}]. For a QD in an immersed 
(or embedded) geometry, it can be written in terms of the QD Green's function as
\begin{equation}
\tau(\omega,T)=\Gamma \Im [G_{d}(\omega,T)] ,
\label{Transm}
\end{equation}
where ${\Gamma}_{\gamma}=\pi V^{2} \rho^c(\mu)$, with $\rho^c_{\sigma_r}(\mu)$ 
being the spin-orbit dependent leads' DOS at the Fermi energy, 
and $\Im$ indicates the imaginary part. 

\section{Spin-orbit effects over the density of states}
\label{sec5}

In Kondo physics, the hybridization function at the Fermi energy, 
${\Gamma}_{\gamma}$, is an important quantity. In all the calculations that follow, 
we employed ${\Gamma}_{0}=0.007$~\cite{note2}, in units of ${D}_0$, 
which is the half-width of the conduction band in the absence of SOC~\cite{note0}, and  
the chemical potential is always located at $\mu=0$. 
In addition, we will vary SOC in the interval $0.0 \leq |\gamma| \leq 0.5$, 
which will make $\Gamma_{\gamma}$ vary in accordance with Eq.~\eqref{eq:Deltaren}. 

\begin{figure}[tbh]
\begin{center}
\includegraphics[width=0.4\textwidth]{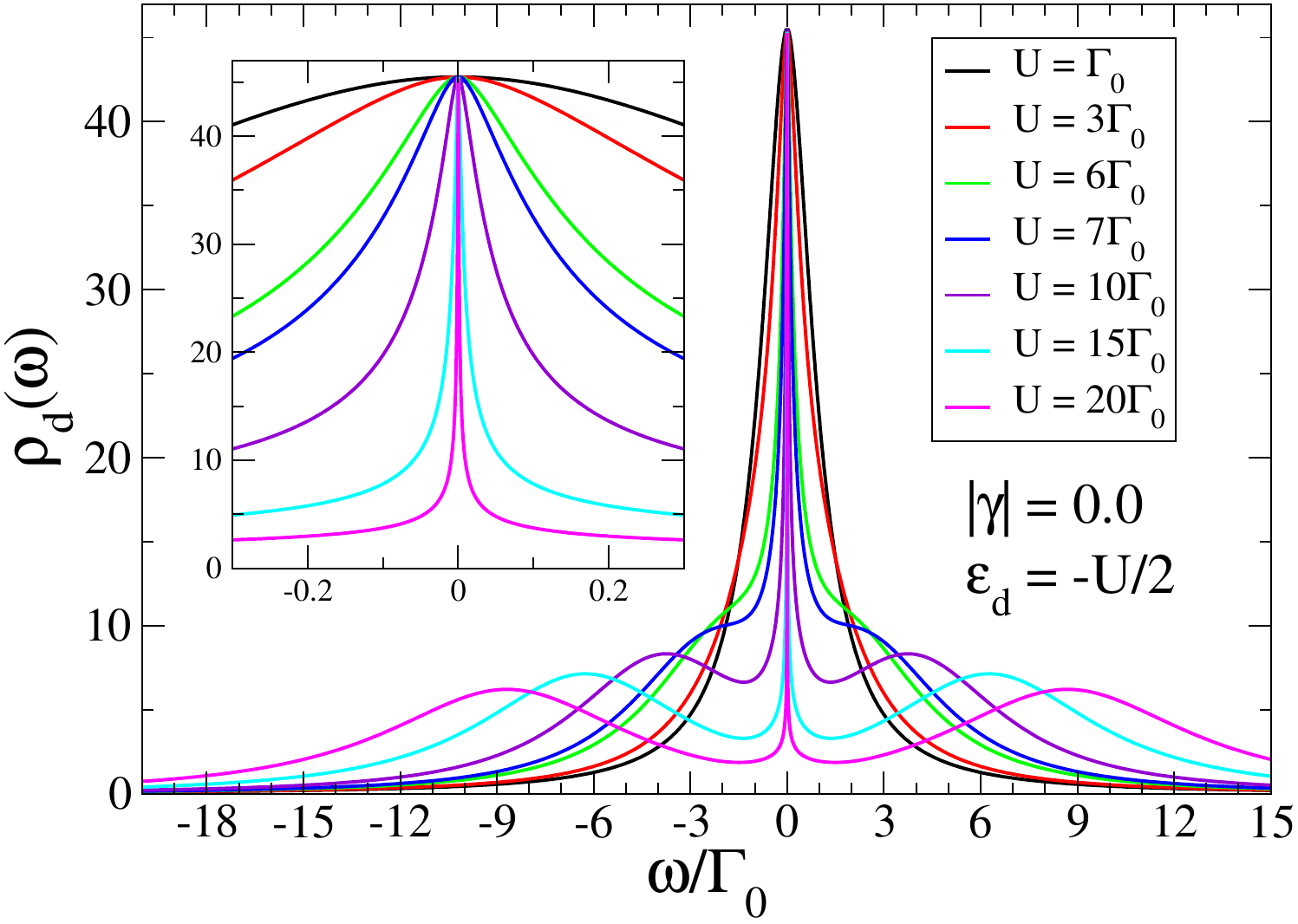}
	\caption {QD's LDOS $\rho_d(\omega)$ for different values of the 
	electronic correlation $U$ and $|\gamma|=0.0$, showing the formation of the Kondo 
	peak, starting from $U=\Gamma_{0}$, when the system is inside the intermediate valence regime, 
	to $U=20.0\Gamma_{0}$, when the system is deep inside the Kondo regime. 
	Inset: Details of the formation of the Kondo peak. Note that all curves 
	are at the PHS point $\epsilon_d = \nicefrac{-U}{2}$.}
\label{fig02}
\end{center}
\end{figure}

In Fig.~\ref{fig02}, we show the QD's LDOS  
$\rho_{d}(\omega)$ at the particle-hole symmetric (PHS) point ($\epsilon_d = -U/2$) 
of the SIAM for several different values of the Coulomb repulsion $\Gamma_0 \leq U \leq 20.0 \Gamma_0$, 
and for vanishing SOC, $|\gamma|=0.0$. For this range of variation of $\nicefrac{U}{\Gamma_0}$, 
it is well know that the system passes from an intermediate valence regime (for the smaller 
values of $\nicefrac{U}{\Gamma_0}$) to the Kondo regime (for the larger values of 
$\nicefrac{U}{\Gamma_0}$). The passage from the former to the latter is a 
crossover, thus it does not occur for an specific value of $\nicefrac{U}{\Gamma_0}$. 
Nonetheless, the results show the gradual formation of the 
Kondo peak as the correlation $U$ is varied from $U=\Gamma_{0}$ to 
$U=20.0\Gamma_{0}$, where the system is already deep into the Kondo regime 
[very low Kondo temperature ${T}_{K\gamma}$---see Eq.~\eqref{eq:hald1}]. For $U=\Gamma_{0}$, we have a 
broad peak centered around the chemical potential, located in $\omega=0$ (black curve); 
in addition, the two symmetric Hubbard satellite peaks characteristic of 
the PHS point cannot be discerned. However, as we increase the 
correlation to $U=10.0\Gamma_0$ (purple curve), the Hubbard satellites 
are already well established, but the Kondo peak is only completely formed 
above $U=15.0\Gamma_0$ (cyan curve). In the inset, the formation 
of the Kondo peak, as the electronic correlation increases, is shown 
more clearly. Along this process, the peak diminishes its width, indicating 
the establishment of the Kondo regime characterized by a Kondo temperature 
$T_{K\gamma}$, which is proportional to the width of the Kondo peak, thus, the narrower 
the peak, the lower is the Kondo temperature and the deeper is the system into the 
Kondo regime. 
\begin{figure}[tbh]
\begin{center}
\includegraphics[width=0.4\textwidth]{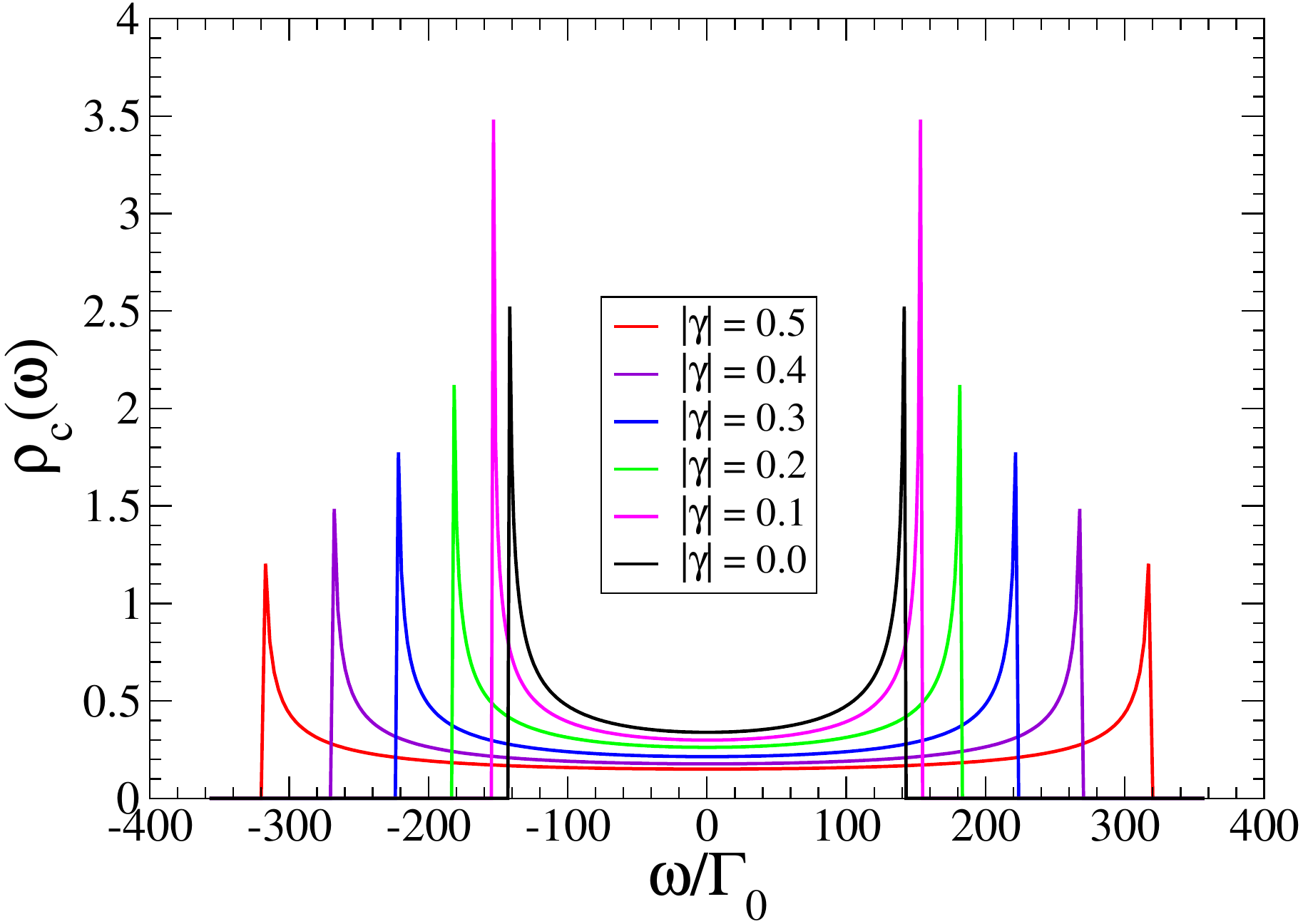}
\caption {Density of states of the 1D conducting leads $\rho_c(\omega)$ for different values of SOC, 
	$0.0 \leq |\gamma| \leq 0.5$. Notice the sizable broadening of the band, 
	as well as the decrease of the DOS at the Fermi energy.}
\label{fig03}
\end{center}
\end{figure}
In Fig.~\ref{fig03}, we show the DOS of the conducting leads 
$\rho_{c}(\omega)$, corresponding to different values of SOC, $0.0 \leq |\gamma| \leq 0.5$. 
The main effects of the SOC is to produce a broadening of the band, and, as a 
consequence, a decrease of the DOS at the chemical potential 
$\mu=0$, which, see Eq.~\eqref{eq:Deltaren}, results in the decrease of the 
value of the SOC-renormalized hybridization function at the Fermi energy.

\begin{figure}[tbh]
\begin{center}
\includegraphics[width=0.4\textwidth]{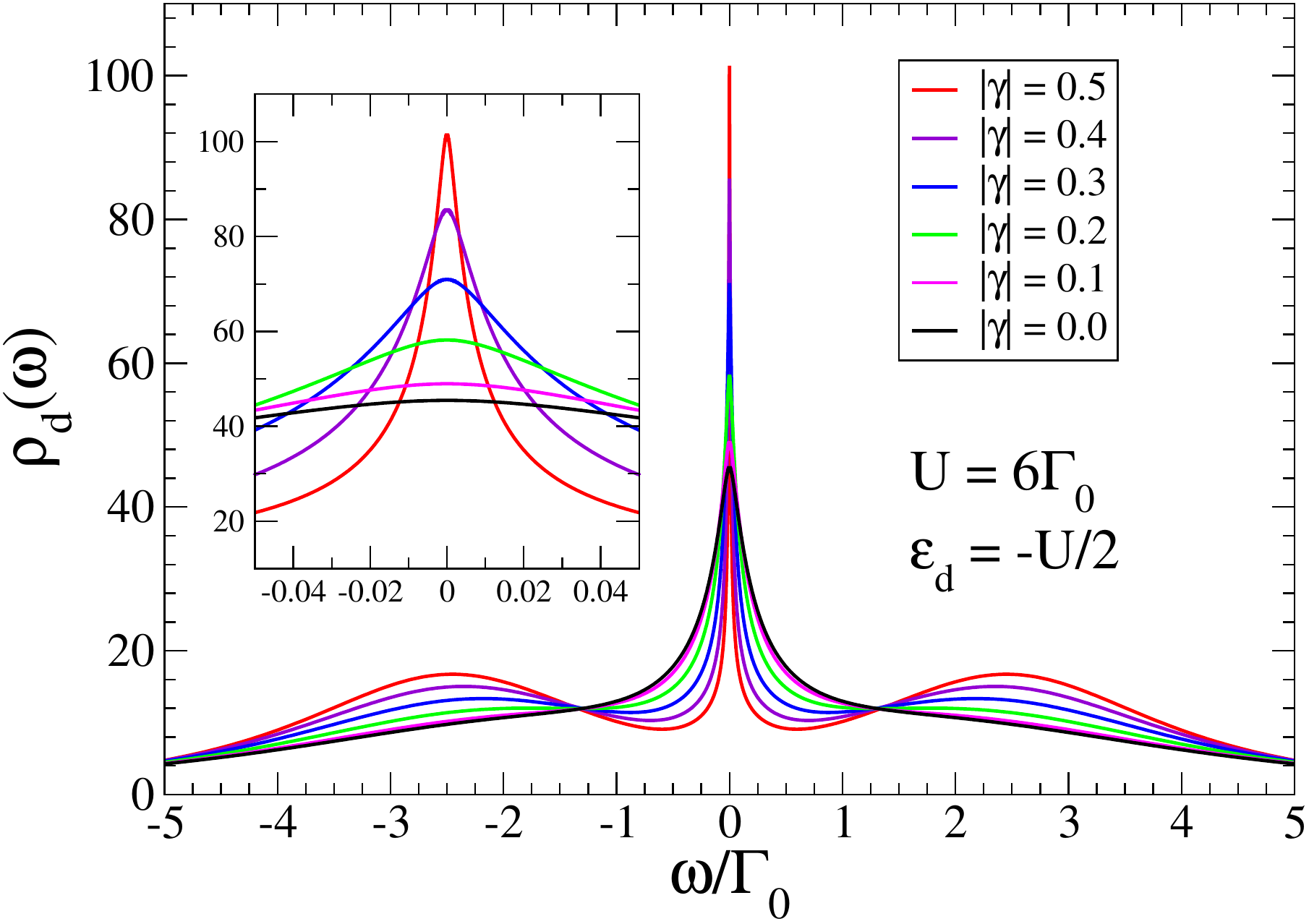}
\caption {QD's LDOS $\rho_{d}(\omega)$, for different SOC values, $0.0 \leq |\gamma| \leq 0.5$, 
	at the PHS point, for $U=6.0\Gamma_0$. Note that the black curve 
	($|\gamma|=0.0$) corresponds to the green curve in Fig.~\ref{fig02}, 
	thus inside the intermediate valence regime, clearly showing that SOC drives the system 
	deep into the Kondo regime. Notice the very well formed Kondo peak for $|\gamma|=0.5$ 
	(red curve). Inset: zoom close to $\omega=0$, showing details of the 
	evolution of the Kondo peak.}
\label{fig04}
\end{center}
\end{figure}

\begin{figure}[tbh]
\begin{center}
\includegraphics[width=0.35\textwidth]{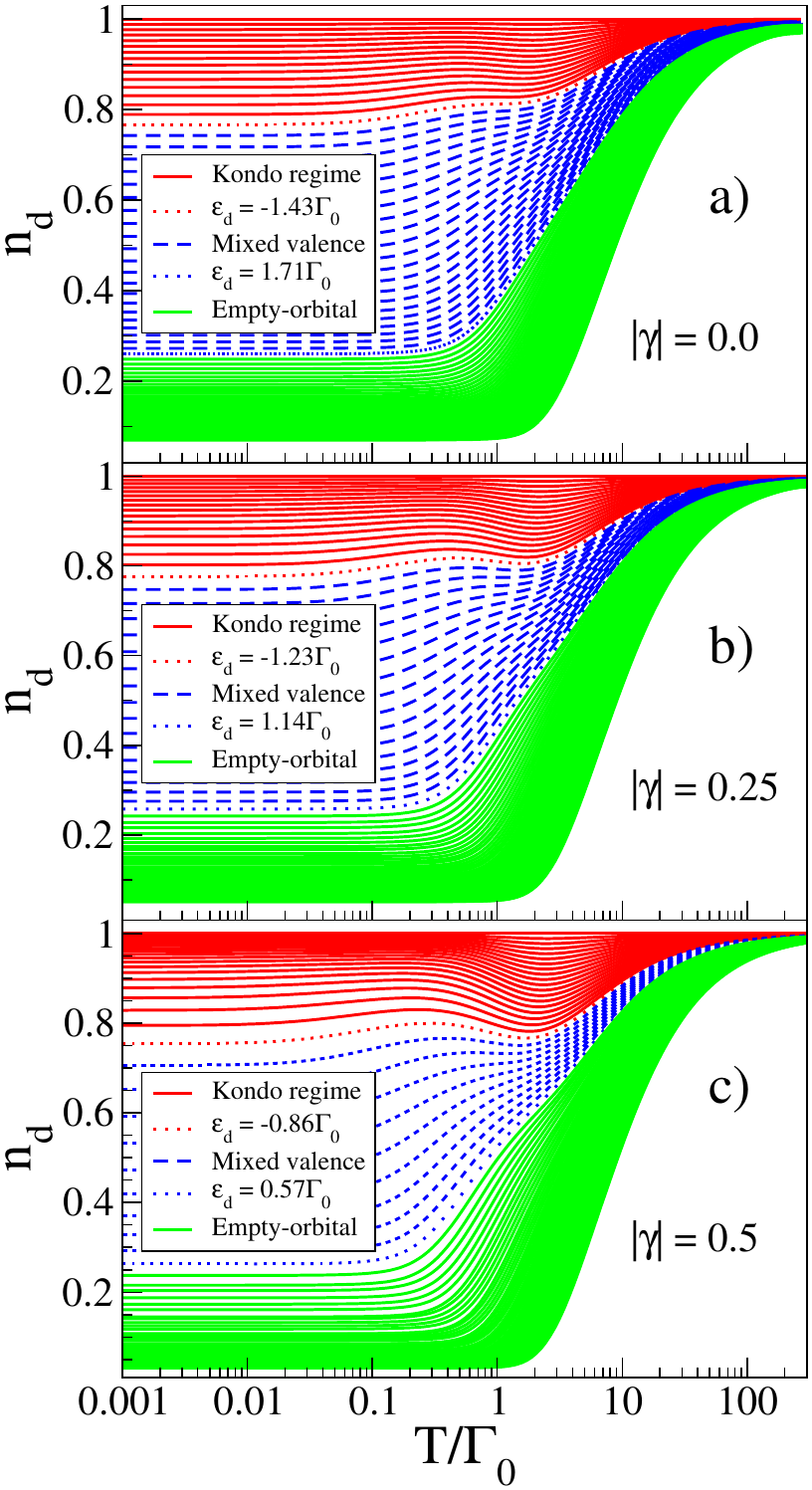}
	\caption {QD occupation-number map. Panels (a), (b) and (c), 
	for $|\gamma|=0.0$, $0.25$, and $0.5$, respectively, indicate 
	the temperature variation of $n_d$ for varying $\epsilon_d$ 
	in the interval $-U/2 \leq \epsilon_d \leq 8.71\Gamma_0$. The definition 
	of the different regimes follows Ref.~\cite{costi2010thermoelectric}, i.e., 
	Kondo (red), $|n_d - 1|_{T \approx 0} \leq 0.25$, intermediate-valence 
	(blue), $|n_d - 0.5|_{T \approx 0} \leq 0.25$, empty-orbital (green), 
	$|n_d|_{T \approx 0}\leq 0.25$. The dotted curves, with corresponding 
	$\epsilon_d$ values indicated in the legends, demarcate the 
	crossover from one regime to the next. All results for $U=7.0\Gamma_0$}
\label{Occupation}
\end{center}
\end{figure}

In Fig.~\ref{fig04}, we plot the QD's LDOS $\rho_{d}(\omega)$, 
for different values of  $|\gamma|$, for a PHS situation. 
We do all the calculations for $U=6.0\Gamma_0$, thus, at $|\gamma|=0.0$, not deep 
into the Kondo regime (see green curve in Fig.~\ref{fig02}). However, 
by the evolution of $\rho_{d}(\omega)$, due to the increase of SOC from 
$|\gamma|=0$ to $|\gamma|=0.5$, it can be clearly seen that 
the increase of SOC drives the system deep into the Kondo regime.  
Indeed, the height of the Kondo peak increases while its width decreases, 
indicating a lowering of the Kondo temperature. 
This striking effect is directly related to the decreasing value of the SOC-renormalized 
hybridization function at the Fermi level, since, according to Friedel's sum rule [Eq.~\eqref{fried}], 
this should cause an increase of $\rho_{d} (\mu=0)$ and, according to Eq.~\eqref{eq:hald1}, 
a decrease of the Kondo temperature, with an accompanying reduction of the  
Kondo-peak half-width. It is also possible to discern a slight increase in the separation 
between the satellite Hubbard peaks, pointing to an increase of the effective Hubbard on-site repulsion, 
which accounts for an increase in the electronic correlations. 

\section{Thermoelectric properties maps}
\label{sec6}

In this section, we plot the TE properties 
for different values of $\epsilon_d$, from the Kondo to the 
empty-orbital regime, as a function of $T/\Gamma_0$, for different values of SOC, 
$|\gamma| = 0.0$, $0.25$, and $0.5$. For all the results in this section, 
we consider the electronic correlation $U=7.0\Gamma_{0}$, and, to characterize 
the different regimes of the system (at low temperature), we follow the definitions 
in Ref.~\cite{costi2010thermoelectric}, viz., 
(i) $n_d$ values in the interval $|n_d - 1|_{T \approx 0} \leq 0.25$ (red curves) 
correspond to the Kondo regime, (ii) $|n_d - 0.5|_{T \approx 0} \leq 0.25$ (blue curves) 
correspond to the mixed-valence regime, (iii) $|n_d|_{T \approx 0}\leq 0.25$ (green curves) 
correspond to the empty-orbital regime. The borders between different regimes 
occur as crossovers. Although these regime definitions are more  
appropriate to the low temperature region, we extend them to the higher 
temperature regions as well. 

In panels (a), (b), and (c) of Fig.~\ref{Occupation}, we plot the QD 
occupation number $n_d$ as a function of temperature for different 
values of $\epsilon_d$ ($-U/2 \leq \epsilon_{d} \leq 8.71\Gamma_0$). 
Panels (a), (b), and (c) are for $|\gamma|=0.0$, 
$0.25$ and $0.5$, respectively. The $\epsilon_{d}$ values shown 
in the legend represent the values at which, according to the definitions above, there is a crossover 
between different regimes, indicated by dotted curves. 
By comparing different panels, it is clear that SOC affects the overall spread of each region. 
Indeed, as $|\gamma|$ increases from $0.0$ to $0.5$, the empty-orbital and Kondo regions 
expand, at the expense of the mixed-valence region. This makes sense, as the decrease of the 
hybridization between the QD and the conduction band, caused by SOC, should enhance spin fluctuations 
(enhancing Kondo and empty-orbital) at the expense of charge fluctuations (weakening intermediate 
valence). As we shall see next, this will be reflected in the results for the TE properties. 

\begin{figure}[tbh]
  \centering
  \includegraphics[width=0.35\textwidth]{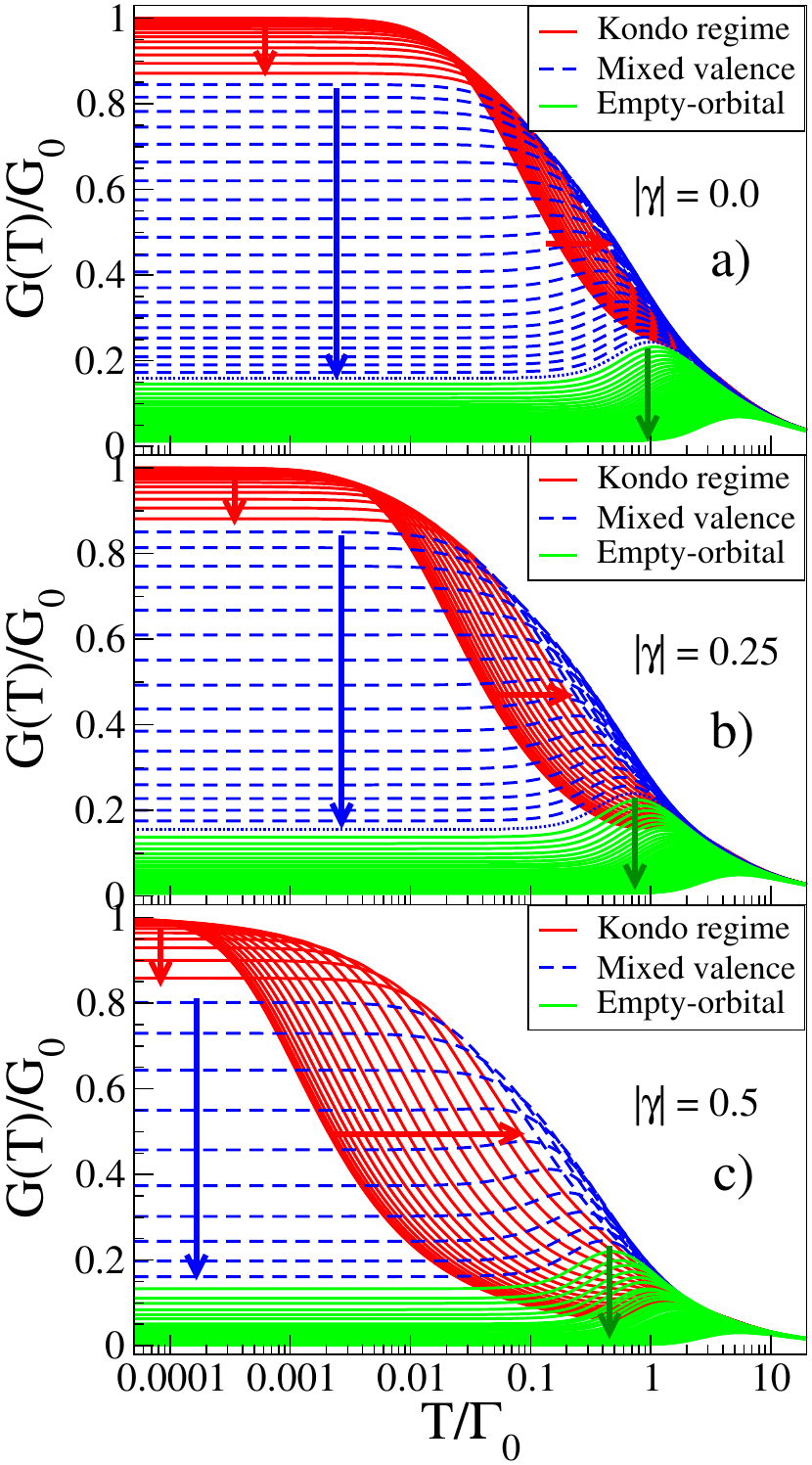}
	\caption{Same parameters as in Fig.~\ref{Occupation}, but now showing the electrical 
	conductance $G(T)$ (in units of the quantum of conductance $G_0$) 
	as a function of temperature. 
	The arrows indicate the direction of increasing values of 
	$\epsilon_d$, where the values of $\epsilon_d$ for each curve are the same 
	as in Fig.~\ref{Occupation}. The temperature is in units of $\Gamma_0$ 
	and $U=7.0\Gamma_0$.}
\label{Conduti}
\end{figure}

In panels (a), (b), and (c) in Fig.~\ref{Conduti}, we have similar plots to the ones in 
Fig.~\ref{Occupation}, but this time for the electrical conductance $\nicefrac{G(T)}{G_0}$, 
as a function of temperature, where $G_0=\nicefrac{2e^2}{h}$ is the quantum of conductance 
(taking spin into account). The arrows indicate the direction of increasing values of 
$\epsilon_d$, where the values of $\epsilon_d$ for each curve are the same as in Fig.~\ref{Occupation}. 
Following Ref.~\cite{Goldhaber1998PRL}, the Kondo temperature, $T_{K\gamma}$, 
can be calculated, from each curve in all three panels, 
by computing the temperature value where the electrical conductance attains 
$G(T_{K\gamma})=\nicefrac{G_{0}}{2}=\nicefrac{e^{2}}{h}$. 
By using that criterion to define $T_{K\gamma}$, 
it is easy to see that the average $T_{K\gamma}$ 
of the red curves (Kondo regime, as defined by Costi \emph{et al.}~\cite{costi2010thermoelectric}) 
in panel (c) is more than an order of magnitude lower than 
the average $T_{K\gamma}$ in panel (a). Taking in account the 
universally accepted concept that, the lower is $T_{K\gamma}$, 
the deeper we are into the Kondo regime, leads us to 
assert that an increase in SOC drives the SIAM deeper into 
the Kondo regime, as already observed through the LDOS results in Fig.~\ref{fig04}.

\begin{figure}[tbh]
  \centering
  \includegraphics[width=0.35\textwidth]{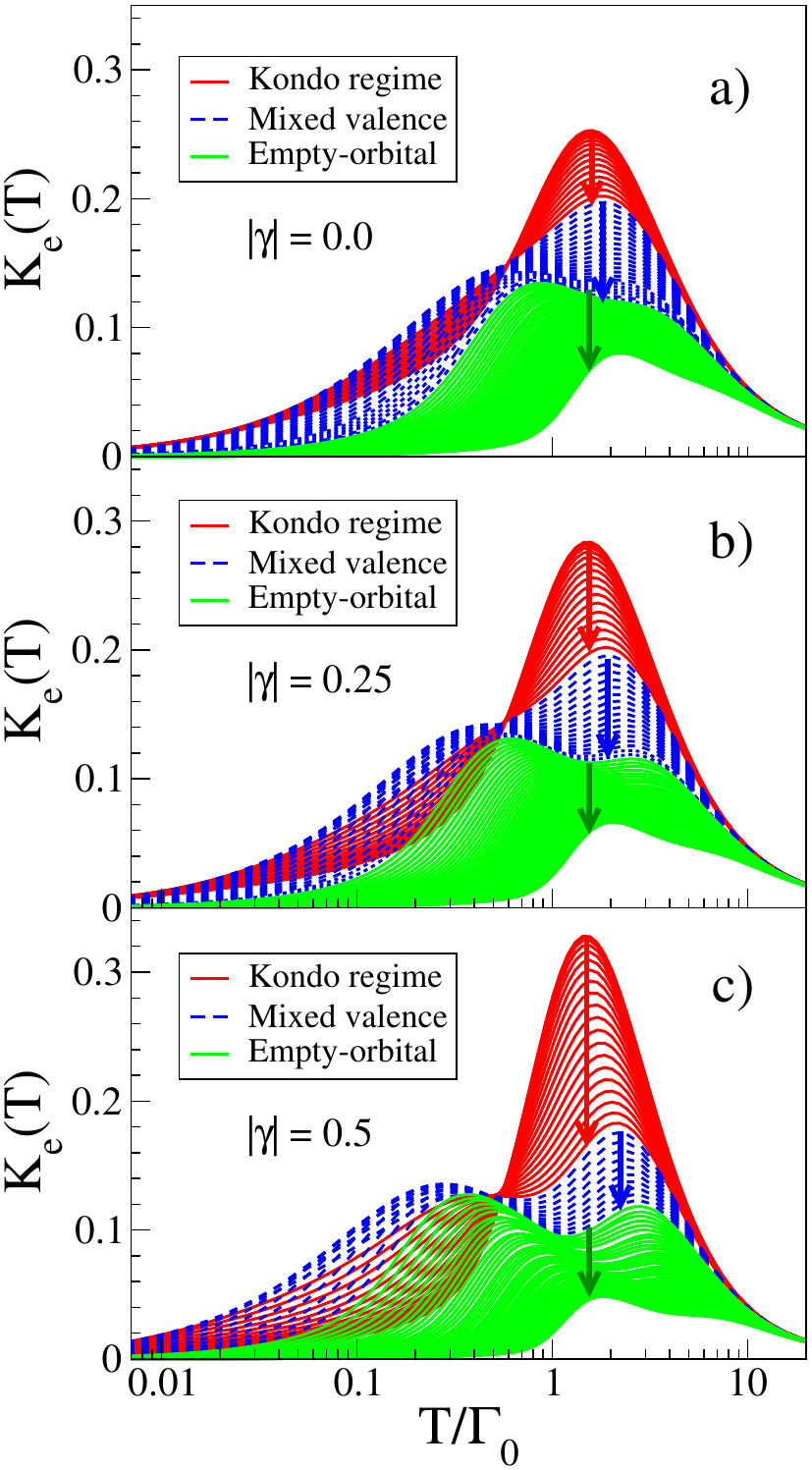}
	\caption{Same parameters as in Figs.~\ref{Occupation} and \ref{Conduti}, 
	but now showing the electronic contribution to the thermal conductance $K_e(T)$ as a 
	function of temperature, in units of $\Gamma_0$. The arrows indicate the direction 
	of increasing values of $\epsilon_d$.}
  \label{Thermcond}
\end{figure}
In panels (a), (b), and (c) of Fig.~\ref{Thermcond}, we have similar plots to the ones in 
Figs.~\ref{Occupation} and \ref{Conduti}, but this time for the thermal conductance $K_e$, 
as a function of temperature, in units of $\Gamma_0$. The arrows have the same meaning as 
in Fig.~\ref{Conduti}. Comparing 
the three panels in Fig.~\ref{Thermcond}, we observe again the SOC's tendency to reduce the intermediate 
valence region and to increase the Kondo and the empty-orbital regions. All three
panels (a), (b), and (c), for $|\gamma|=0.0$, $0.25$, and $0.5$, respectively, 
exhibit a crossing point, slightly below $T=\Gamma_{0}$ (and weakly dependent on $\gamma$). 
This crossing point appears as the convergence 
of all the red and blue curves to a very narrow window interval at 
$T\approx\Gamma_{0}$~\cite{costi2010thermoelectric}. It is interesting to note that 
the width of this window becomes increasingly narrower as $\gamma$ increases, basically 
collapsing to a single point for $|\gamma|=0.5$. These crossing points are characteristic 
signatures of strongly correlated systems, like it was observed 
for the specific heat in the Hubbard model in Ref.~\cite{vollhardt1997characteristic}. 

\begin{figure}[tbh]
  \centering
  \includegraphics[width=0.35\textwidth]{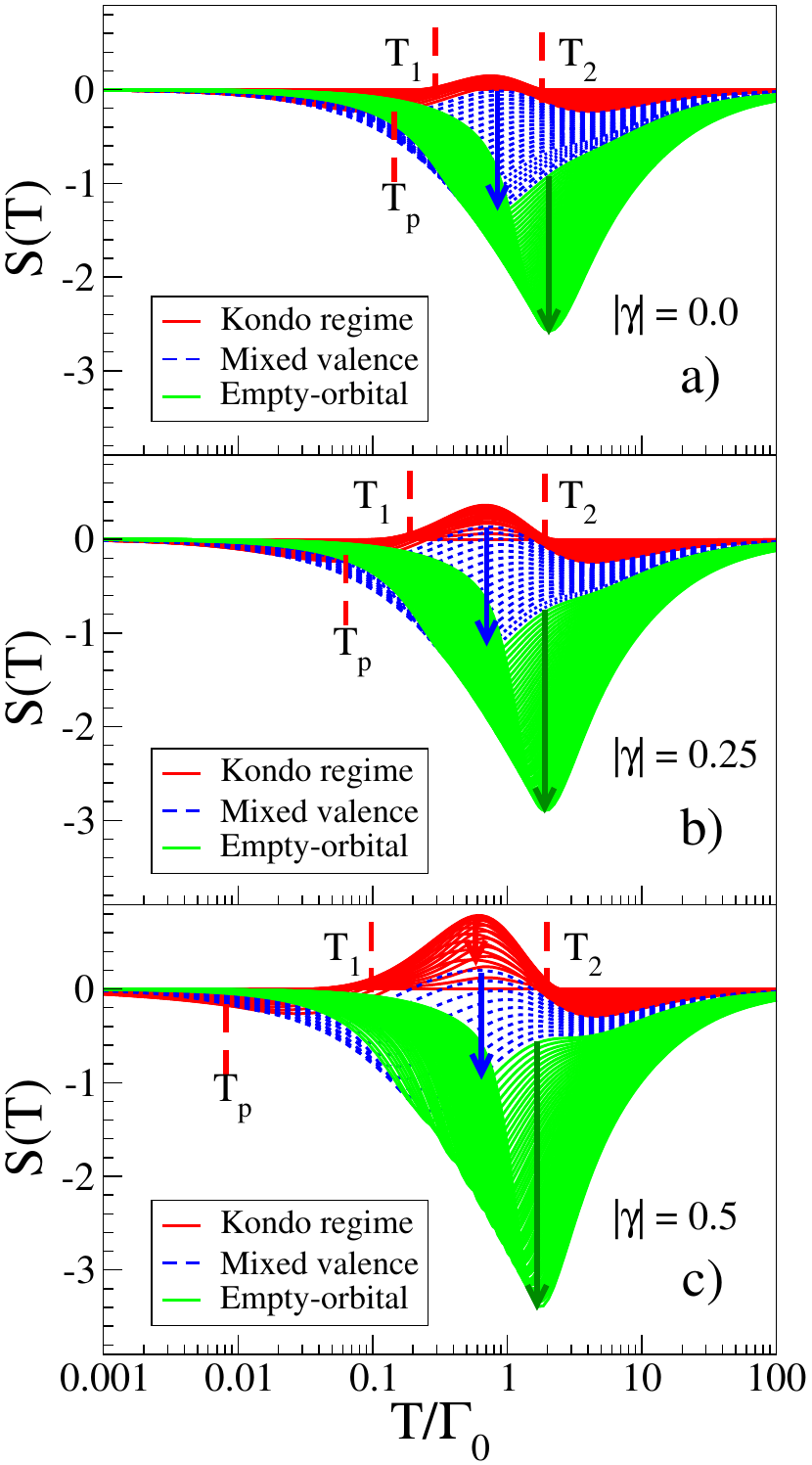}
	\caption{Same parameters as in Figs.~\ref{Occupation}, \ref{Conduti} and \ref{Thermcond}, 
	but now showing the thermopower $S(T)$ as a function of temperature. 
	Notice the sizable change in the interval of variation of S(T) (increase in the 
	maximum and minimum values) as a function of $\gamma$. This will be relevant 
	to the TE figure of merit results in Fig.~\ref{fig13}. Again, the arrows indicate the direction 
	of increasing values of $\epsilon_d$.}
  \label{Thermopmap}
\end{figure}
In panels (a), (b), and (c) of Fig.~\ref{Thermopmap}, we show 
plots similar to the ones in Figs.~\ref{Occupation}, \ref{Conduti}, 
and \ref{Thermcond}, but this time for the thermopower $S$, as a 
function of temperature. There are three peaks in the Kondo regime (red curves), viz., 
two minima satellite peaks located at left and right of a maximum 
central peak located at $T \approx \Gamma_0$, which is inside the interval 
$[T_1: T_2]$. Temperatures $T_1$ and $T_2$ represent energy scales associated to the 
Kondo regime that characterize the changes in who 
are the $S(T)$ heat carriers, from electrons to holes to electrons, from left to right. 
In the PHS point, i.e., $\epsilon_d=\nicefrac{-U}{2}$, $S(T) = 0$, however, 
away from the PHS point, $S(T)$ acquires a temperature dependence. Comparing 
the three panels, for $|\gamma|=0.0$, $0.25$ and $0.5$, when the system is not 
in the PHS point, there is a strong increase 
in the height of the maximum Kondo-related peak (red curves) and in the depth of the 
minimum empty-orbital-related  
peak (green curves), as $|\gamma|$ increases from $0.0$ to $0.5$, which is the most striking characteristic 
of the thermopower shown here. That will contribute to the sizable $ZT$ increase seen in Fig.~\ref{fig13}(c), 
at finite $\gamma$, when compared to zero-SOC [Fig.~\ref{fig13}(a)]. 

\section{Universality under SOC}
\label{sec7}

In this section, we present a study of the universal behavior of the electrical and 
thermal conductances, as well as of the thermopower, as a function of 
temperature, for different values of $\epsilon_{d}$, 
for $|\gamma|=0$, and how this universal behavior changes for varying SOC.

\begin{figure}[htb]
  \centering
 \includegraphics[width=0.4\textwidth]{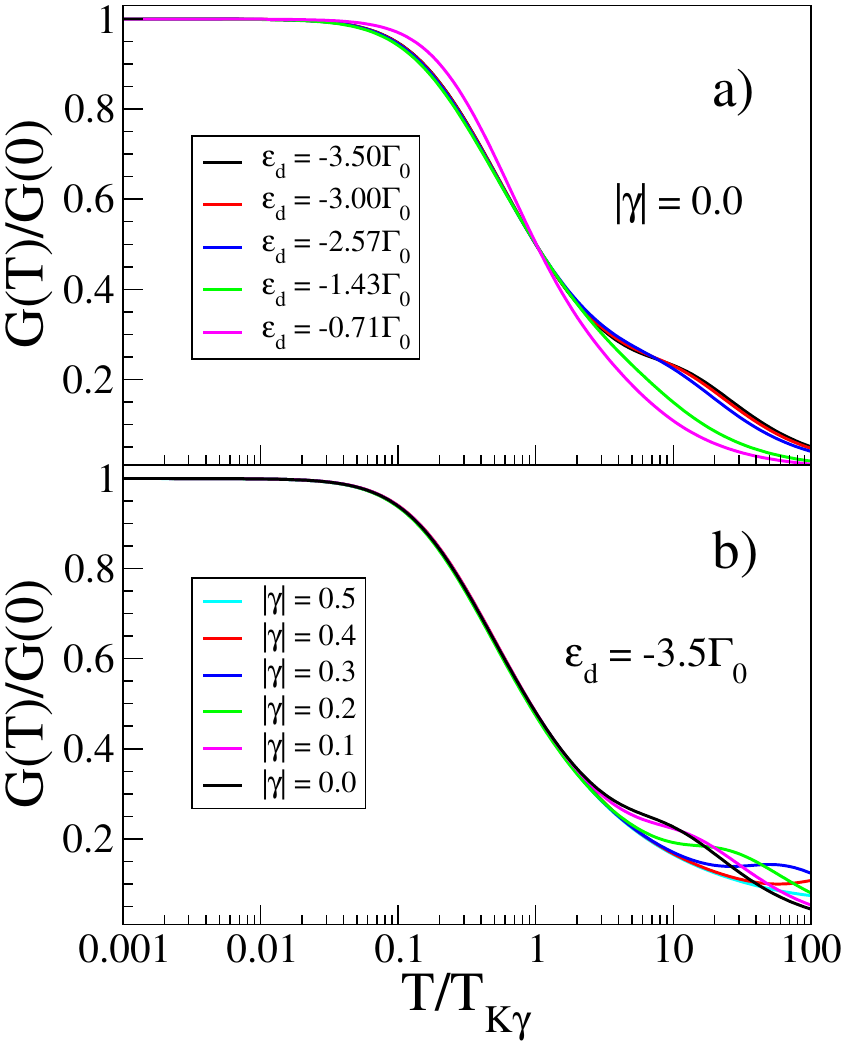}
	\caption{(a) Universal behavior of the electrical conductance $\nicefrac{G(T)}{G(0)}$ 
		for different values of $\epsilon_d$ and 
		$|\gamma|=0.0$ as a function of the scaled temperature 
		$\nicefrac{T}{T_{k\gamma}}$. The universality 
		occurs inside the Kondo regime, since the magenta curve, which does not collapse, 
		falls already inside the intermediate valence regime. 
		(b) Universal behavior of $\nicefrac{G(T)}{G(0)}$, 
	as a function of the scaled temperature $\nicefrac{T}{T_{K\gamma}}$, where all curves for 
	different SOC ($0.0 \leq |\gamma| \leq 0.5$) collapse into a single curve $f(T/T_{K\gamma})$ 
	for temperatures up to $T \gtrsim T_{K\gamma}$.}
 \label{Conduc}
\end{figure}
In Fig.~\ref{Conduc}(a), we plot the electrical conductance $\nicefrac{G(T)}{G(0)}$ 
as a function of the scaled temperature $\nicefrac{T}{T_{K\gamma}}$, for several $\epsilon_{d}$ values,  
for $|\gamma|=0.0$i, where $G(0)$ is given by Eq.~(10) in Ref.~\cite{costi2010thermoelectric}. 
As expected, the curves for the first four values of $\epsilon_d$, which fall 
inside the Kondo regime, collapse into a single curve. On the other hand, the cyan curve, for 
$\epsilon_d=-0.71\Gamma_0$, which is inside the intermediate valence regime [see Fig.~\ref{Occupation}(a)], 
does not collapse into the other curves. 
The situation is similar if we stay at the PHS point and 
vary $\gamma$. In Fig.~\ref{Conduc}(b) we plot $\nicefrac{G(T)}{G(0)}$ in 
the PHS point, $\epsilon_d = \nicefrac{-U}{2}$, as a function of the scaled temperature $\nicefrac{T}{T_{K\gamma}}$, 
for different values of SOC, $0.0 \leq |\gamma| \leq 0.5$.  In the Kondo regime, the electrical 
conductance presents a universal character: 
$\nicefrac{G(T)}{G(0)} = f(\nicefrac{T}{T_{K\gamma}})$, with a functional 
form that is independent of SOC. Note that the larger is $|\gamma|$, the 
further above $T_{K\gamma}$ remains the invariance of $f(\nicefrac{T}{T_{K\gamma}})$ with $\gamma$. 

We just saw that a quite interesting characteristic of electronic transport through QDs 
is the universal behavior in the Kondo regime when the temperature is 
scaled by a characteristic temperature, such as $T_{K\gamma}$ for $G(T)$, as just shown above, 
or by $T^{\theta}_{K\gamma}$ for $K_{e}(T)$. The temperature $T^{\theta}_{K\gamma}$ is the equivalent 
of $T_{K\gamma}$ for $K_{e}(T)$ and can be computed using the Wiedemann-Franz law, Eq.~\eqref{LL},  
$\nicefrac{K_e}{T} \approx L_o \times WF \times G(T)$, 
being defined by the relation~\cite{costi2010thermoelectric}
\begin{equation}
	\frac{K_e(T=T_{K\gamma}^{\theta})}{T^{\theta}_{K\gamma}} = \frac{\alpha}{2} ,
	 \label{Kscale}
\end{equation}
 where  $\alpha$ is obtained through
\begin{equation}
  \alpha = \lim_{T\rightarrow 0} \frac{K_e(T)}{T} .	
\label{Alph}	
	\end{equation}
\begin{figure}[htb]
  \centering
 \includegraphics[width=0.4\textwidth]{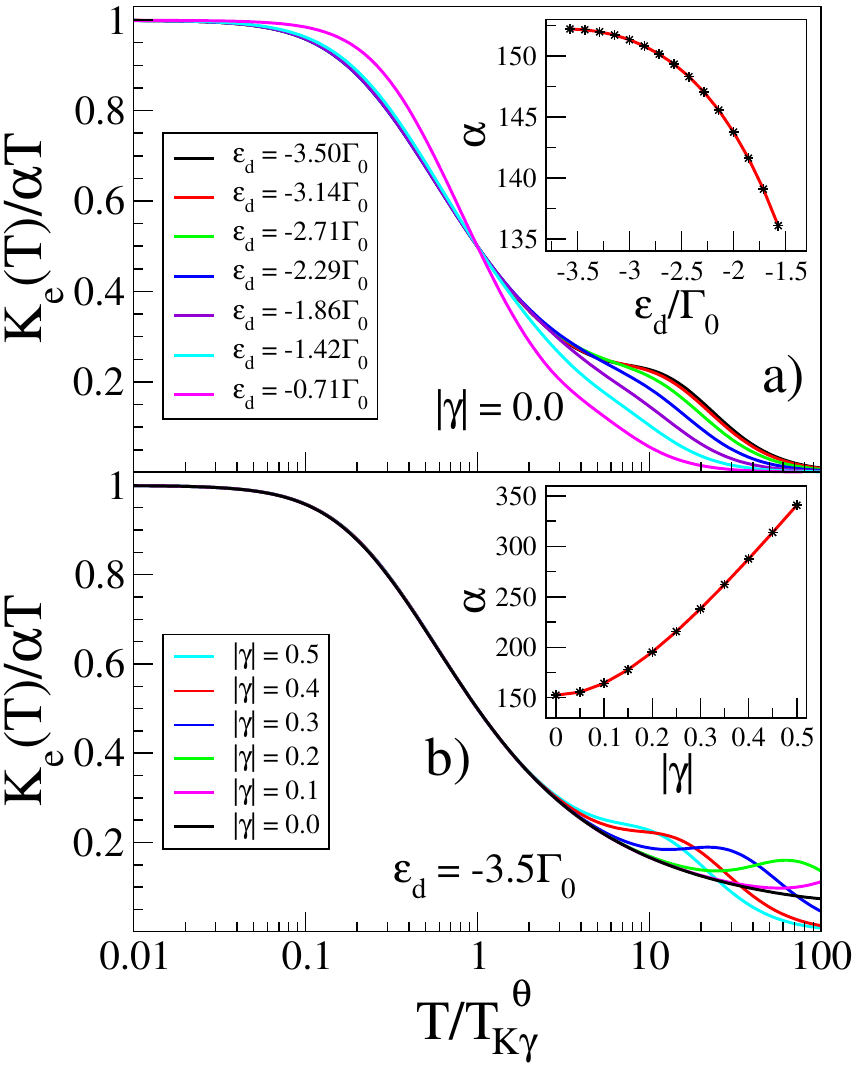}
	\caption{(a) Universal behavior of the thermal conductance $\nicefrac{K_{e}(T)}{\alpha T}$  
	as a function of the scaled temperature $\nicefrac{T}{T^{\theta}_{K\gamma}}$, for several 
	values of $\epsilon_d$ (inside the Kondo regime), for $|\gamma|=0.0$. The single curve inside 
	the intermediate valence regime ($\epsilon_d=-0.71\Gamma_0$, magenta curve) does not collapse 
	into the Kondo regime universality function. 
	(b) Universal behavior of $\nicefrac{K_{e}(T)}{\alpha T}$, 
	as a function of the scaled temperature $\nicefrac{T}{T^{\theta}_{K\gamma}}$, where all curves for 
	different SOC ($0.0 \leq |\gamma| \leq 0.5$) collapse into a single curve $g(T/T^{\theta}_{K\gamma})$. 
	In both insets we show the values of $\alpha$ that produce the collapse.}
 \label{ConducK1}
\end{figure}
In Fig.~\ref{ConducK1}(a), we plot the thermal conductance $\nicefrac{K_{e}(T)}{\alpha T}$
as a function of the scaled temperature $\nicefrac{T}{T^{\theta}_{K\gamma}}$, 
for several $\nicefrac{-U}{2} \leq \epsilon_{d} \leq -0.71 \Gamma_0$ and
$|\gamma|=0.0$. In the inset to panel (a), we plot the rescaling parameter 
$\alpha$ as a function of $\epsilon_{d}$, in $\Gamma_{0}$ units. 
It is clear from the results that the rescaling by $T^{\theta}_{K\gamma}$ and $\alpha$ 
collapses all the $K_e(T)$ curves, for different $\epsilon_d$, for $T \lesssim T^{\theta}_{K\gamma}$, 
onto a single universal curve. The exception, as in the case of the electric conductance, was for 
$\epsilon_d=-0.71$ (magenta curve), which is inside the intermediate valence regime. 
In Fig.~\ref{ConducK1}(b) we plot $\nicefrac{K_{e}(T)}{\alpha T}$ at the PHS point, for different values 
of $|\gamma|$. In the Kondo regime, the thermal conductance thus presents a universal character: 
$\nicefrac{K_{e}(T)}{\alpha T} = g(\nicefrac{T}{T^{\theta}_{K\gamma}})$, showing its invariance with SOC. 
In addition, in the PHS point, the thermal conductance obeys, 
by construction [see Eqs.~\eqref{Kscale} and \eqref{Alph}],  
$\nicefrac{K_{e}(T)}{\alpha T} =1.0$, at low temperatures. In the inset to panel (b), we plot the 
rescaling parameter $\alpha$ as a function of $|\gamma|$.

\begin{figure}[tbh]
  \centering
  \includegraphics[width=0.4\textwidth]{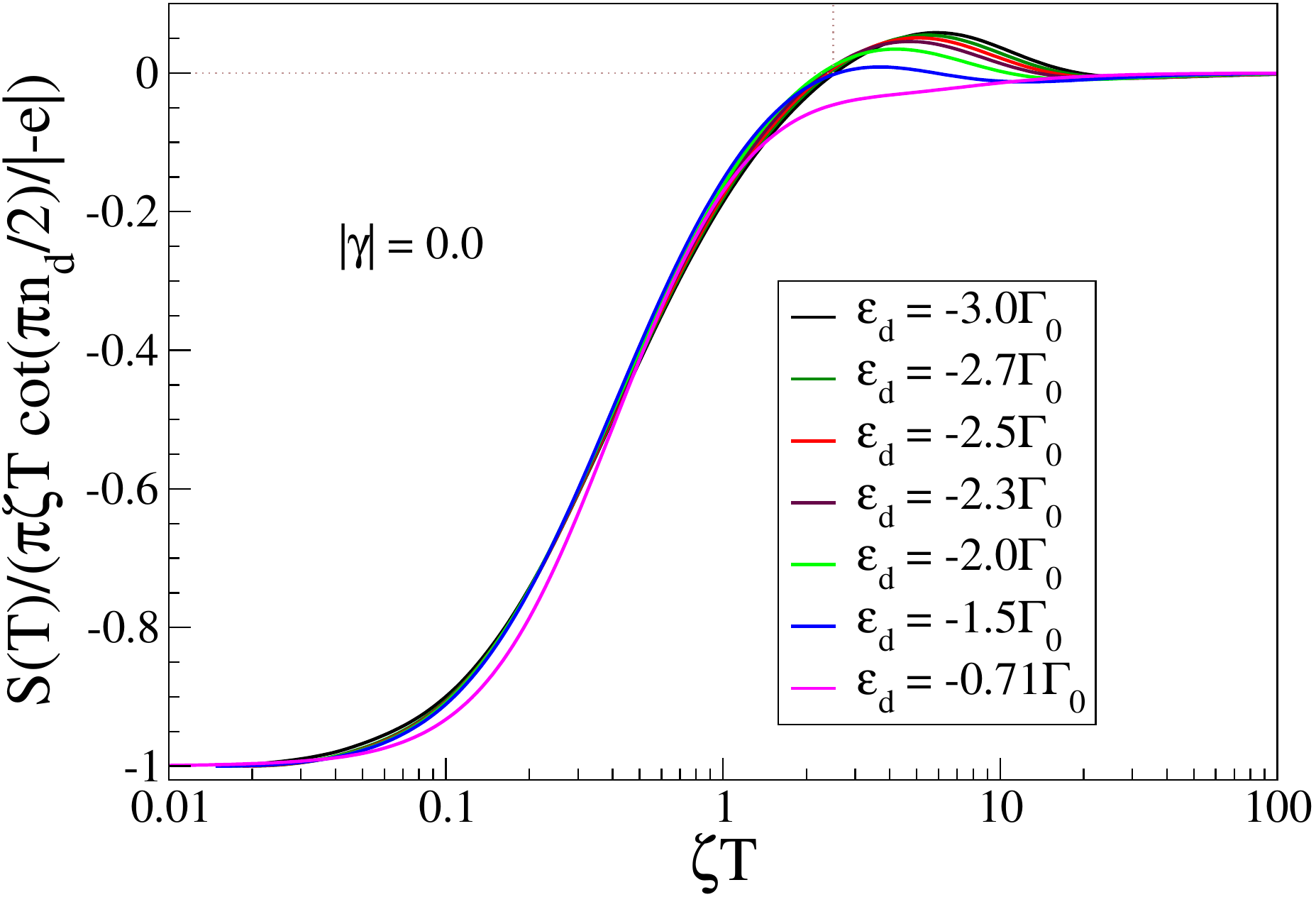}
	\caption{Temperature dependence of the thermopower $S(T)$, plotted 
	in units of $\nicefrac{\pi \zeta T \cot(\nicefrac{\pi n_{d}}{2})}{e}$ 
	for $-3.0\Gamma_0\leq \epsilon_d \leq -0.71\Gamma_0$ and $|\gamma|=0.0$. 
	Universality is achieved for $T \lesssim \zeta T$. As it happened for 
	the electric and thermal conductances, the curve for the first value 
	inside the intermediate valence regime ($\epsilon_d=-0.71$, magenta curve) 
	does not collapse into the universal curve.}
  \label{Thermop1}
\end{figure}

\begin{figure}[tbh]
\begin{center}
\includegraphics[width=0.45\textwidth]{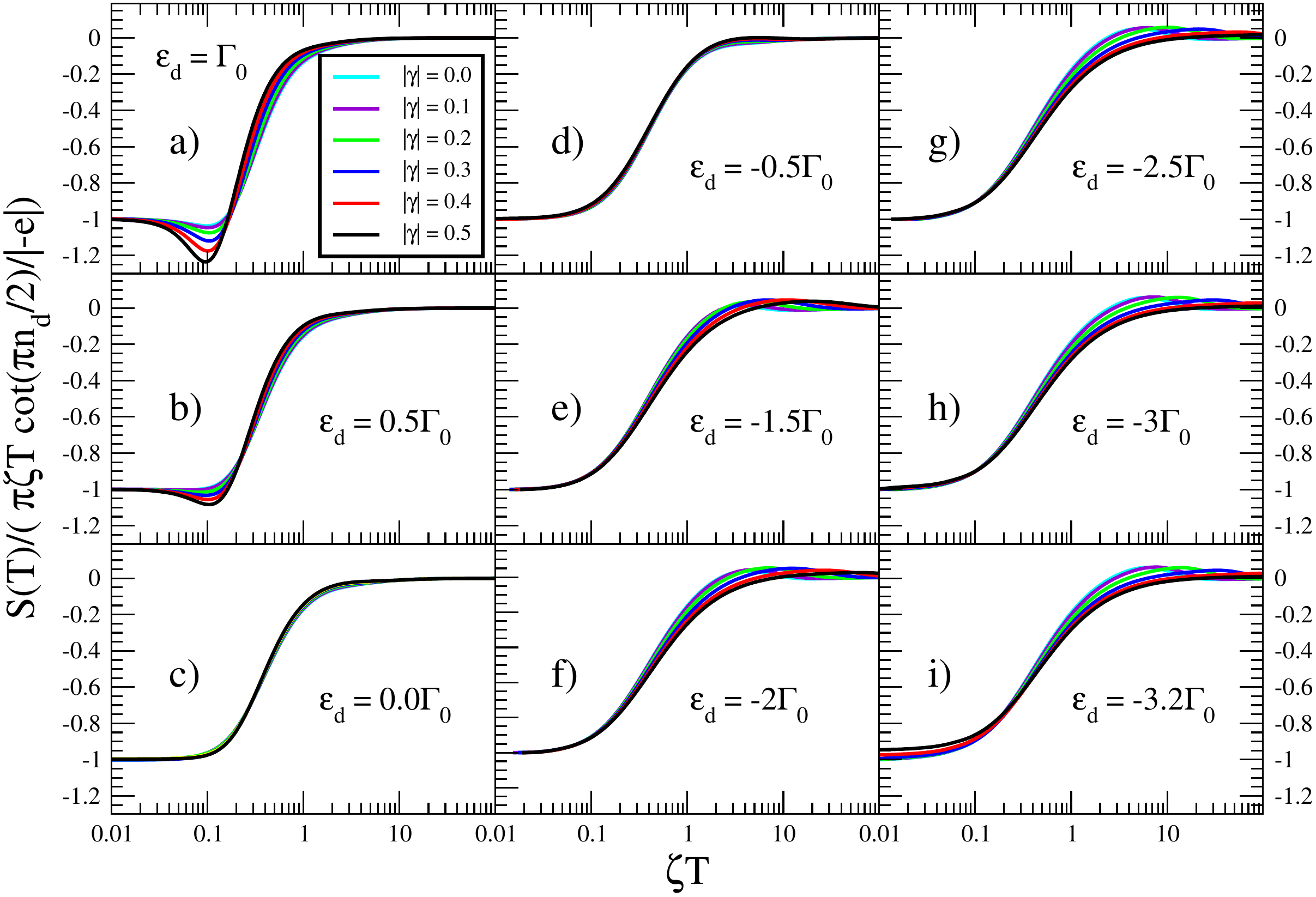}
	\caption {(a) to (i): Comparison of thermopower universality in the 
	Kondo and intermediate valence regimes. Each panel contains the 
	scaled thermopower (for $0.0 \leq |\gamma| \leq 0.5$) 
	for different values of $\epsilon_d$. Notice how the universality 
	is more complete in the intermediate valence regime. Indeed, 
	panels (c) and (d), for $\epsilon_d=0.0\Gamma_0$ and $-0.5\Gamma_0$, 
	present the more complete collapse of the thermopower results 
	for different values of $\gamma$. These two values of 
	$\epsilon_d$ are deep into the intermediate valence regime, 
	for all values of $\gamma$ [see Fig.~\ref{Occupation}(c)].}
\label{uniS}
\end{center}
\end{figure}
As pointed out by Costi \emph{et al.}~\cite{costi2010thermoelectric}, in the Fermi liquid 
regime~\cite{Costi1993}, $\nicefrac{S(T)}{T}$, for a range of different values of $\epsilon_d$ in the 
Kondo regime, scales as
\begin{equation}
	\frac{S(T)}{T}=-\frac{\pi \zeta}{e} \cot(\frac{\pi n_d}{2}) ,
\label{STscale} 
\end{equation}
where $-e$ is the electron charge, and the factor $\zeta$ can be obtained from the numerical value of 
$\lim_{T\rightarrow 0}|S(T)/T|$ and the occupation number $n_{d}$. 

As done for the electric and thermal conductances (Figs.~\ref{Conduc} and \ref{ConducK1}), 
we will employ this procedure to rescale the temperature dependence of $S(T)$ to check the  
universality for varying $\epsilon_d$ (at $|\gamma|=0.0$) and for varying $\gamma$ at fixed 
$\epsilon_d$. In Fig.~\ref{Thermop1}, we plot the thermopower $S(T)$, 
in units of $\nicefrac{\pi \zeta T \cot(\nicefrac{\pi n_{d}}{2})}{e}$, as a function 
of the scaled temperature $\zeta T$, for several $\epsilon_{d}$ 
and $|\gamma|=0.0$. In agreement with what we obtained for the electric and thermal conductances,  
$S(T)$ attains universality if we stay inside the Kondo regime, i.e., 
$ -3.0\Gamma_0\leq \epsilon_d \leq -1.5\Gamma_0$. For $\epsilon_d=-0.71$ (magenta curve) the 
universality is lost. In addition, since the sign of $S(T)$ is determined by the charge of the heat 
carriers ($S(T)>0 \leftrightarrow$~holes, and $S(T)<0 \leftrightarrow$~electrons), 
for temperatures below $\zeta T\simeq 2.0$ (see black dotted lines in Fig.~\ref{Thermop1}), 
the carriers are electrons, and, in a region above $\zeta T\simeq 2.0$ the carriers are 
holes. In the limit of high temperatures, $S(T) \rightarrow 0$. 

Something curious, however, occurs when we analyze the universality at fixed $\epsilon_d$ and 
$0.0 \leq |\gamma| \leq 0.5$. As shown in Fig.~\ref{uniS}, where panels (a) to (i) show the scaling of 
$S(T)$ for different values of $\epsilon_d$ in the interval $-3.2\Gamma_0 \leq \epsilon_d \leq \Gamma_0$ 
(spanning the Kondo and intermediate valence regimes), the universality is achieved only deep into 
the intermediate valence regime (panels (c) and (d), for $\epsilon_d=-0.5\Gamma_0$ and 
$0.0\Gamma_0$, respectively). This is in contrast to what was observed for the electric 
and thermal conductances [Figs.~\ref{Conduc}(b) and \ref{ConducK1}(b)], where the universality 
was observed inside the Kondo regime.

\begin{figure}[tbh]
  \centering
  \includegraphics[width=0.4\textwidth]{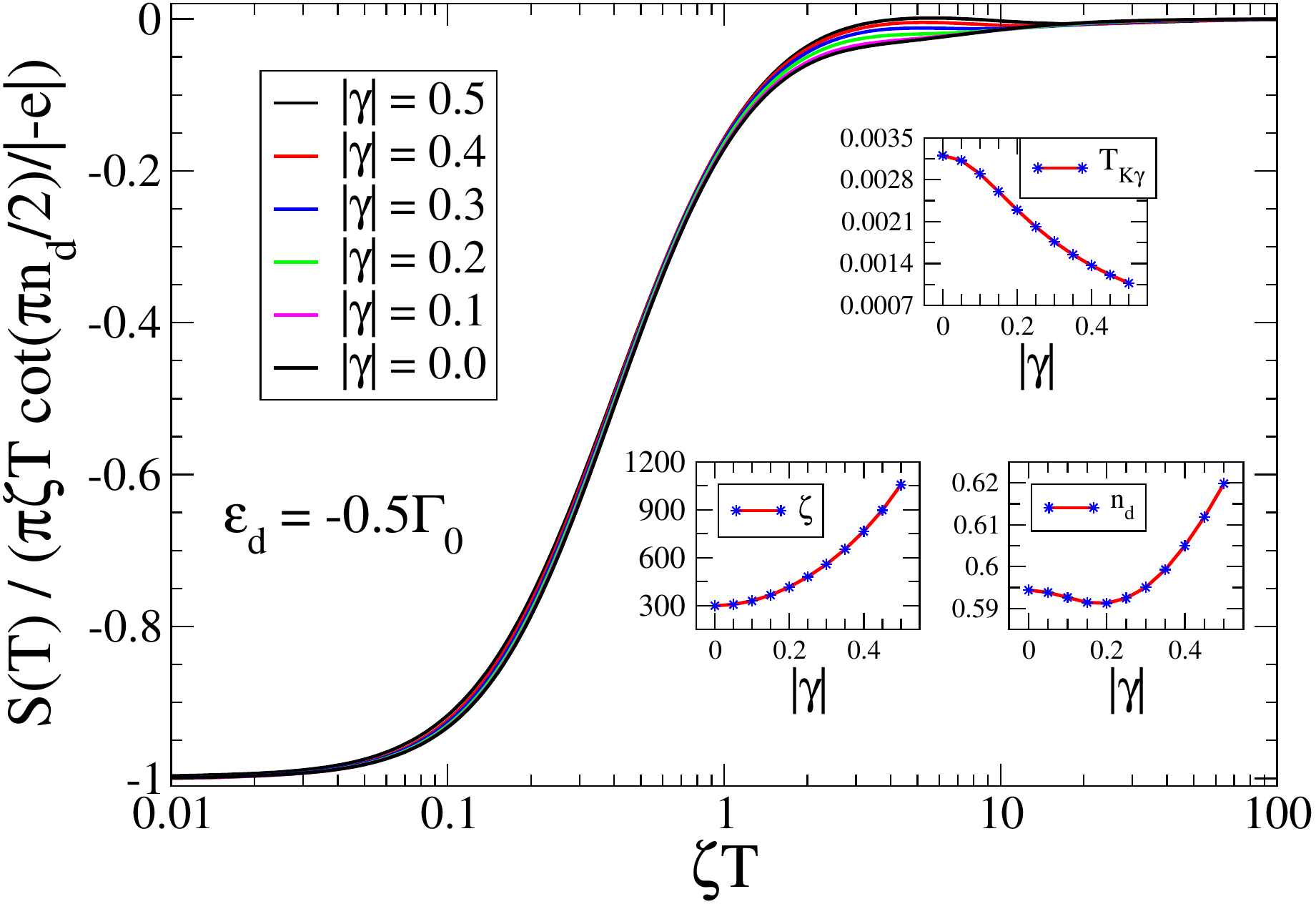}
	\caption{ Same as in Fig.~\ref{Thermop1}, but for 
	$0.0\leq |\gamma|\leq 0.5$ and $\epsilon_d=-0.5\Gamma_0$. 
	Top inset: $T_{K\gamma}$ as a function of $\gamma$; bottom-left inset: 
	$\zeta$ as a function of $\gamma$; bottom-right inset: QD occupation $n_d$ 
	as a function of $\gamma$.}
  \label{Thermop2}
\end{figure}

In Fig.~\ref{Thermop2}, we re-plot Fig.~\ref{uniS}(d) ($S(T)$ for $\epsilon_d=-0.5\Gamma_0$) 
to study the variation of $T_{K\gamma}$ with $\gamma$ (top inset), the 
dependence of the Fermi liquid parameter $\zeta$ with $\gamma$ (bottom-left inset), 
and the dependence of the QD occupancy $n_d$ with $\gamma$ (bottom-right inset). 
As expected, since the increase in $\gamma$ moves the system in the Kondo regime direction, 
we see that there is a non-monotonic increase in $n_d$ as $\gamma$ increases (bottom-right inset), 
while, as expected too, $T_K$ decreases with $\gamma$ (top inset). 
In addition, there is a corresponding increase in $\zeta$ with $\gamma$ (bottom-left inset). 

Thus, we have analyzed two types of universalities for the quantities $G(T)$, 
$K_e(T)$, and $S(T)$: (i) zero-SOC and varying $\epsilon_d$, for which we found that 
there is universality for $G(T)$, $K_e(T)$, and $S(T)$ in the Kondo regime 
[see Figs.~\ref{Conduc}(a),~\ref{ConducK1}(a), and \ref{Thermop1}]; 
(ii) fixed $\epsilon_d$ and varying $\gamma$, for which both 
$G(T)$ and $K_e(T)$ show universality in the Kondo regime [see Figs.~\ref{Conduc}(b) 
and \ref{ConducK1}(b)], while, unexpectedly, $S(T)$ shows universality in the intermediate 
valence regime (Fig.~\ref{uniS}). We are not completely sure why this is so.

\begin{figure}[tbh]
\begin{center}
\includegraphics[width=0.4\textwidth]{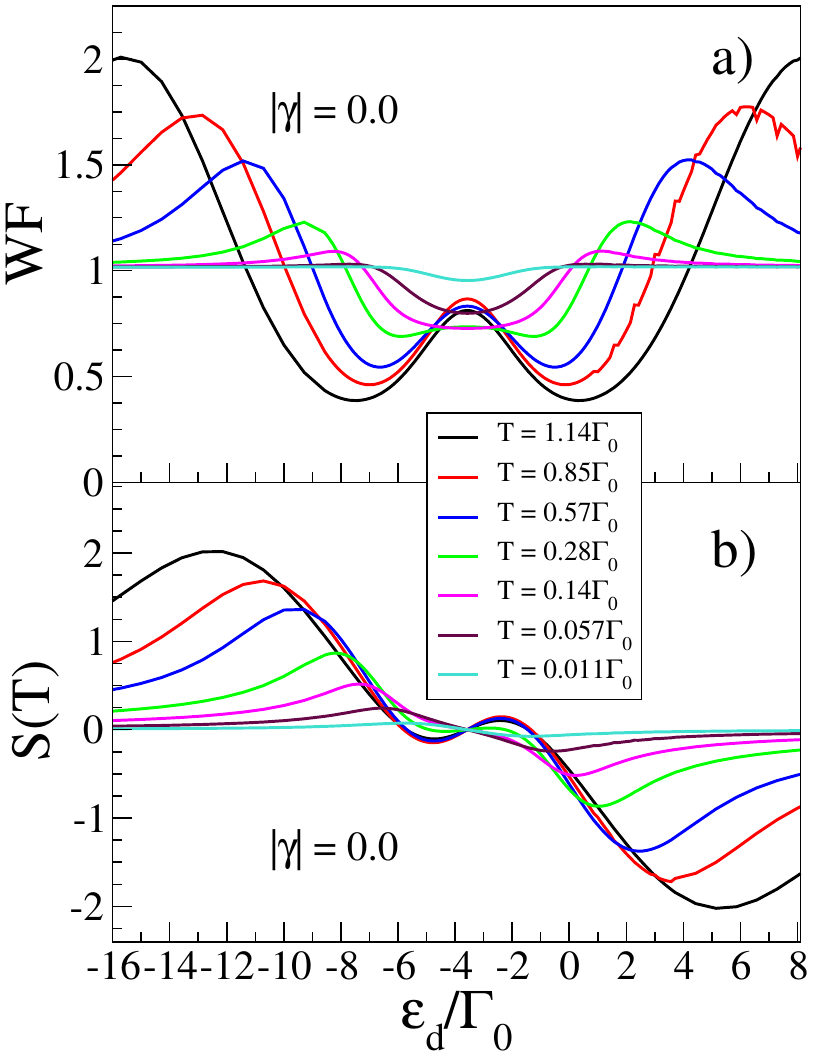}
	\caption {(a) Wiedemann-Franz law (in units of the Lorenz number, $L_o$) and (b) Thermopower, 
	as a function of $\epsilon_d$, for several values of temperature (in units of $\Gamma_0$), 
	for $U=7.0\Gamma_0$ and $|\gamma|=0.0$.}
\label{WF}
\end{center}
\end{figure}
In panels (a) and (b) in Fig.~\ref{WF}, we show the Wiedemann-Franz 
law, in units of the Lorenz number $L_{o}$, and the thermopower, 
respectively, as a function of 
$\epsilon_{d}$ (in units of $\Gamma_{0}$), at various temperature 
values (also in units of $\Gamma_{0}$), for $|\gamma|=0$ and $U=7.0\Gamma_0$.  
At the lowest temperature ($T=0.011\Gamma_0$, cyan curve), the Wiedemann-Franz law is satisfied, 
aside from a small region around the PHS point ($\epsilon_{d}=-3.5 \Gamma_{0}$), where $WF \lesssim 1$. 
As the temperature increases, the width of this region increases, as 
well as the departure of $WF$ from $1$. In addition, two broad peaks appear farther 
away from the PHS point (on the left and right of it), whose violation of the 
Wiedemann-Franz law (now, $WF > 1$) becomes more severe, as the temperature increases. 
In addition, the maxima of the left and right 
peaks gradually move away from the PHS point with increasing temperature. 

\begin{figure}[tbh]
\begin{center}
\includegraphics[width=0.4\textwidth]{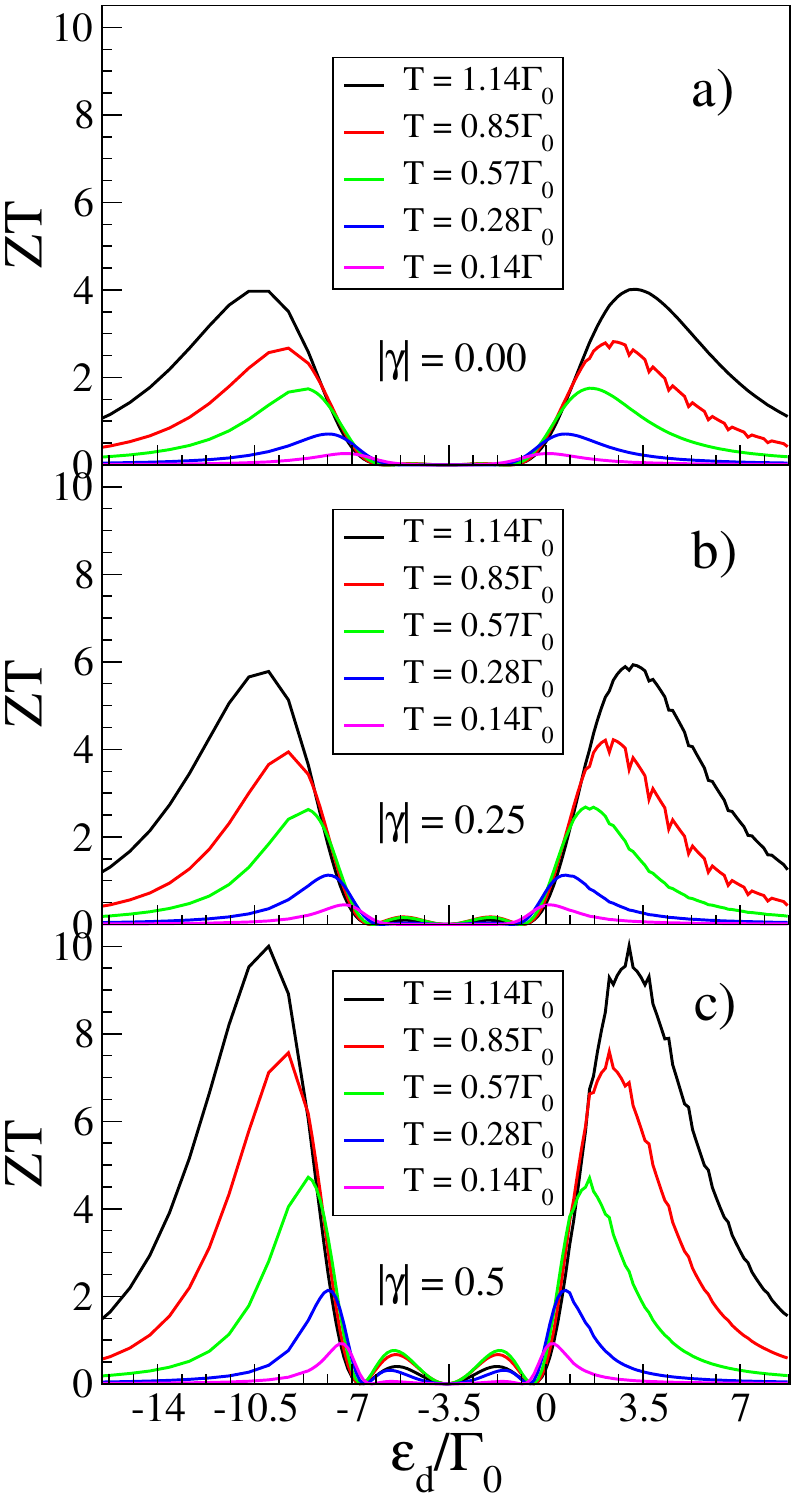}
\caption {Dimensionless TE figure o merit $ZT$ as a function of $\epsilon_{d}$, 
	for several values of temperature (in units of ${\Gamma_0}$) and $U=7.0\Gamma_0$. 
	Panels (a), (b), and (c) are for $|\gamma|=0.0$, $0.25$, and $0.5$, respectively.}
\label{fig13}
\end{center}
\end{figure}

A somewhat similar picture describes the results for $S(T)$ in Fig.~\ref{WF}(b), 
with the difference that now S(T) is odd in relation to the PHS point. In addition, 
left and right broad peaks emerge away from the PHS point, similarly located and with similar temperature 
dependence as the ones shown for $WF$ in 
Fig.~\ref{WF}(a). As a consequence, given that $ZT=\nicefrac{S^2}{WF}$, and since $S \gtrsim WF$ 
at and around those broad peaks, this determines the relatively high values attained by $ZT$ 
in the peaks region, as shown in Fig.~\ref{fig13}(a), for $|\gamma|=0.0$ and several temperatures. 

In panels (b) and (c) in Fig.~\ref{fig13}, we show the dimensionless TE figure 
of merit $ZT$ as a function of $\epsilon_{d}$, at various temperatures, 
for finite SOC, $|\gamma|=0.25$ and $0.5$, respectively. 
When compared to Fig.~\ref{fig13}(a), for $|\gamma|=0.0$, we observe a sizable 
enhancement of $ZT$ with SOC, which results from the increase of $S(T)$ with SOC, as 
indicated in the $S(T)$ maps in Fig.~\ref{Thermopmap}. We notice that, compared to 
the $|\gamma|=0.0$ maximum $ZT \approx 4.0$ results in Fig.~\ref{fig13}(a) ($T=1.14 \Gamma_0$), 
the $ZT \approx 10.0$ obtained for $|\gamma|=0.5$, for the same temperature, represents 
an improvement in ZT of $\approx 2.5$ times. 

Finally, we should note that, as previously mentioned, 
in the calculation of these $ZT$ results, we do not consider any phononic contribution, 
which tends to compete with the electronic contribution to decrease the $ZT$ values as the temperature 
is increased. However, note that we kept the maximum temperature studied at a low enough 
value that justifies the neglect of phonons. 

 \section{Conclusions and perspectives} 

\label{sec8}

In summary, we have studied the effect of 1D conduction band SOC over 
the TE transport properties of an SET. This was done, using NRG, through 
the calculation of temperature maps of the TE properties. We have shown 
that SOC drives the system deeper into the Kondo regime.  We also showed 
that the Kondo regime universality of thermal and electrical conductances 
is maintained in the presence of SOC. We also show the interesting result 
that $S(T)$, which is universal in the Kondo regime at zero-SOC, presents 
a more universal behavior (for different $\gamma$) in the intermediate 
valence regime, when compared to the Kondo regime. 
More importantly, we have shown that the large increases in 
the thermopower, caused by SOC (see Fig.~\ref{Thermopmap}), translate 
into notable SOC-caused enhancements of the TE figure of merit 
ZT (see Fig.~\ref{fig13}) for an embedded SET coupled to 1D leads. Interesting 
points to consider in future research are (i) how these results would change 
for a side-connected SET; (ii) what is the role played by the
leads dimensionality in the sizable increase of the figure of merit
observed for 1D leads; (iii) the Rashba and Dresselhaus conduction band SOC 
results obtained here can be extended to study two-dimensional (2D) systems, like the surface 
states of the Kondo insulator $SmB_{6}$. Some recent experimental results 
point out that a combination of Rashba- and Dresselhaus-like 
SOC~\cite{Xu2014,Zhu_2016,Li2020,Ryu2021} 
can describe the states around the $X$ point of the Brillouin zone; 
(iv) finally, we would like to study the 
TE properties of an SET embedded in a 2D electron gas at 
the Persistent Spin Helix point ($\alpha=\beta$)~\cite{Bernevig2006}.

\section{Acknowledgments}

We thank CAPES,  CNPq and FAPERJ for the support of this work.
G.~B.~M. acknowledges financial support from the Brazilian agency Conselho Nacional de 
Desenvolvimento Cient\'{\i}fico e Tecnol\'ogico (CNPq), processes 424711/2018-4,  305150/2017-0 and M.~S.~F.  acknowledges financial support from the Brazilian agency Funda\c c\~ao de Amparo a Pesquisa do Estado do Rio de Janeiro, process 210 355/2018.

\bibliography{bibliography}

\begin{thebibliography}{81}
\bibitem [{not({\natexlab{a}})}]{note0}%
  \BibitemOpen
  \href@noop {} {} ({\natexlab{a}}),\ \bibinfo {note} {our unit of energy will
  be the half-bandwidth at zero-SOC, i.e., ${D}_0=1$.}\BibitemShut {Stop}%
\bibitem [{not({\natexlab{b}})}]{note2}%
  \BibitemOpen
  \href@noop {} {} ({\natexlab{b}}),\ \bibinfo {note} {the meaning of this
  statement is the following: the choice of $V$ is such that the
  \emph{zero-SOC} hybridization function at the Fermi energy, ${\Gamma}_0$, is
  $0.007$ [see Eq.~\eqref{eq:Deltaren}]. This same $V$ is used for the
  finite-SOC calculations, resulting in a hybridization that decreases with SOC
  (see Fig.~\ref{fig03}).}\BibitemShut {Stop}%
\bibitem [{not({\natexlab{c}})}]{note1}%
  \BibitemOpen
  \href@noop {} {} ({\natexlab{c}}),\ \bibinfo {note} {we add a subscript $e$
  to the thermal conductivity, $\kappa_e$, to indicate that we refer to energy
  transport by electrons, excluding any contribution from phonons.}\BibitemShut
  {Stop}%
\end{thebibliography}

\end{document}